\documentclass[aps,prd,superscriptaddress,showkeys,showpacs,longbibliography]{revtex4-2}
\usepackage{amsmath,amssymb,amsfonts}
\usepackage{graphicx}
\usepackage{grffile}
\usepackage{siunitx}
\usepackage{bm}
\usepackage[hidelinks]{hyperref}
\usepackage[capitalise]{cleveref}
\usepackage{booktabs}
\usepackage{xcolor}
\usepackage{microtype}
\usepackage{placeins}
\usepackage{subcaption}
\usepackage{multirow}
\usepackage{array}

\begin{document}
\raggedbottom
\title{Compact GridPix Tracking: Test Beam Performance and Detector Response Modelling}
\author{Dhruv Chouhan}
\email{dchouhan@cern.ch}
\affiliation{Rheinische Friedrich-Wilhelms-Universität Bonn, Germany}
\author{Jochen Kaminski}
\affiliation{Rheinische Friedrich-Wilhelms-Universität Bonn, Germany}
\author{Konstantinos Spyrou}
\affiliation{Rheinische Friedrich-Wilhelms-Universität Bonn, Germany}
\author{Matthias Schott}
\affiliation{Rheinische Friedrich-Wilhelms-Universität Bonn, Germany}
\begin{abstract}
\small

We characterise a Timepix-based GridPix detector with a $21\,\mathrm{mm}$ drift length using $2.6\,\mathrm{GeV}$ electrons at ELSA. Tracks entering through the window at several angles are reconstructed as compact $\mu$TPC tracklets, providing position and local direction from one layer. Measured angular widths are approximately $5$--$10\,\mathrm{mrad}$. An empirical response model calibrated to these data augments Geant4 ionisation information with gas-transport and readout effects, reproducing ten of twelve angular widths within approximately $20\%$ and all spatial-residual widths within approximately $12\%$. Application without re-tuning to an independent $5\,\mathrm{mm}$-drift measurement shows larger angular discrepancies. Our response model therefore provides input to feasibility studies of future compact tracking detectors.

\end{abstract}
\maketitle
\section{Introduction}
\label{sec:intro}
GridPix detectors combine a micromesh-based gas-amplification stage with a CMOS pixel readout ASIC, providing fine transverse granularity, single-electron sensitivity, and timing information. They have been developed for TPC tracking~\cite{32chip_part1,32chip_part2} and low-background axion searches~\cite{cast_desch,solar_axions_cast_gridpix}. With a compact drift volume, a particle entering through the window leaves a short tracklet. Pixel positions and arrival times provide both a crossing position and a local direction from a single layer. This configuration is closely related to the $\mu$TPC method in GEM and Micromegas detectors~\cite{micro_tpc_gem,micro_tpc_micromegas}; earlier non-parallel GridPix measurements are reported in Ref.~\cite{gridpix_van_der_graaf}.

Local direction information could support vertex association near an interaction point. One prospective application is a compact tracking station adjacent to an emulsion detector in a future FASER upgrade, where GridPix tracklets could be associated with emulsion vertices. Other possibilities include short-track reconstruction in gaseous active-target neutrino detectors. The present study characterises the detector response but does not establish the feasibility of these applications.

Design studies require a response model that can be used in realistic experimental geometries. Geant4~\cite{Geant4} describes energy deposition and secondary-particle production but does not supply the subsequent electron drift, diffusion, amplification, or pixel timing response. Dedicated tools such as Garfield++~\cite{Schindler2012, GarfieldPPUserGuide}, interfaced with Magboltz~\cite{magboltz}, can describe gas transport microscopically. Here, an empirical response calibrated to test beam data provides a practical alternative for Geant4-based detector studies.

We study a Timepix-based GridPix detector using $2.6\,\mathrm{GeV}$ electrons at ELSA, directed through the window across the full drift gap. Several tilt--rotation settings probe different tracklet orientations. HDBSCAN clustering and Kalman fitting reconstruct both the data and simulated hits, allowing direct comparisons of angular and spatial-residual distributions. The response model first incorporates diffusion, pixelisation, timing, and first-electron arrival. Multiplicity matching accounts empirically for additional hit losses. A probabilistic local-$z$ component then describes longitudinal hit-density variations, and a common effective-smearing term addresses remaining differences in the reconstructed observables. These stages are calibrated with six $21\,\mathrm{mm}$ configurations, then applied without re-tuning to an independent $5\,\mathrm{mm}$ measurement.

Sections~\ref{sec:GridPix_detector} and \ref{sec:test_beam_setup} describe the detector and measurements. Reconstruction is presented in Sec.~\ref{sec:Clustering_Track_Reco}, simulation in Sec.~\ref{sec:MC_Geant4_Simulation}, and response calibration and uncertainties in Sec.~\ref{sec:Detector_Response_Model}. Section~\ref{sec:Results} reports the comparisons and limitations.

\section{GridPix Detector}
\label{sec:GridPix_detector}
 GridPix detectors \cite{gridpix_van_der_graaf} combine micromesh amplification stages with high spatial resolution CMOS pixel readout ASICs. In these detectors, the amplification mesh is directly fabricated on the surface of the ASIC using photolithographic processes and is referred to as the Integrated Grid (InGrid) \cite{ingrid_production}. Owing to the fabrication process, every hole in the mesh is precisely aligned with one pixel of the ASIC. An array of pillars, positioned at dedicated locations between the pixels, is created as part of the fabrication process. The pillars maintain a uniform amplification gap of $50\,\mu\mathrm{m}$\cite{KAMINSKI2017233}. In contrast to traditional strip-based detectors, each pixel of the Timepix ASIC acts as an individual anode\cite{llopart} and integrates charge collection and signal shaping. With a pixel pitch of $55\,\mu\mathrm{m}$, the ASIC comprises a
$256\times256$ array of effectively square pixels, providing an active area of
$14.08\,\mathrm{mm}\times14.08\,\mathrm{mm}$
($1.982\,\mathrm{cm}^2$). The Timepix ASIC allows measurements to be performed in three operational modes: Time-over-Threshold (ToT) for charge measurements, Time-of-Arrival (ToA) for timing measurements, and an additional chessboard mode, in which alternating pixels operate in ToT and ToA modes. The measurements presented in this work were recorded in ToA mode.

The cylindrical body of the detector, with a diameter of $78\,\mathrm{mm}$, forms the drift volume. A GridPix detector with a very similar design and construction was previously employed by CAST~\cite{KRIEGER_timepix_body}. The detector is designed as a modular structure such that the cylindrical component can be replaced, allowing measurements with different drift lengths. The Timepix ASIC, with the grid fabricated on its surface, is mounted on a carrier board inside the readout housing. A field-shaping electrode with an aperture matching the chip dimensions surrounds the ASIC to mitigate field distortions. The cathode, positioned at the top of the drift region, is constructed from FR-4 composite material with a $13 \times 13\,\mathrm{mm}^2$ entrance window. The opening and the surrounding FR-4 are covered by a $100\,\mu\mathrm{m}$-thick aluminium plate.
\section{Test Beam Setup}
\label{sec:test_beam_setup}
Two GridPix detector variants were investigated at the ELSA accelerator with
a $2.6\,\mathrm{GeV}$ electron beam: one with a $21\,\mathrm{mm}$ drift length and one with a $5\,\mathrm{mm}$ drift length. The detector was positioned with its window facing the beam such that each incoming electron traversed the full drift length at selected angular configurations. Both detector variants were operated with an Ar:CO$_2$ (70:30) gas mixture at room temperature and a near-atmospheric pressure of $1.050\,\mathrm{bar}$. A drift field of $500\,\mathrm{V\,cm^{-1}}$ was maintained during both test beam campaigns. Since the detector drift length was varied by changing the detector body, the same Timepix chip was used for both measurements.

The detector readout chain used a Scalable Readout System (SRS) combined with a Front-End Controller (FECv6) and a dedicated adapter board. The FECv6 board provides communication, trigger handling, and data transfer between the Timepix ASIC, the GridPix interconnection motherboard (IMB), and the control computer. The FEC and IMB were connected by two HDMI cables, carrying slow-control and high-speed data signals as well as the low-voltage supply for the IMB. The $2.2\,\mathrm{V}$ bias for the Timepix ASIC was provided by a separate low-voltage supply. High voltages for the grid, cathode, and anode were supplied separately to the IMB through individual SHV connections. Detector operation and data acquisition were controlled using the \texttt{Timepix Operating Software (TOS)} \cite{LUPBERGER_SRS, TOS}.

The chip was operated in ToA mode with a shutter window of $17.25\,\mu\mathrm{s}$. For these measurements, the external trigger was provided by the coincidence of two scintillators placed in front of the detector window. The detector and scintillators were aligned with the incoming beam using a laser such that the beam passed through the centre of the scintillator-overlap region and the centre of the GridPix active area. A veto signal synchronised with the Timepix frame readout inhibited further scintillator triggers until readout completion, preventing overlapping trigger requests within a readout frame.
\begin{figure}[htbp]
    \centering
    \includegraphics[height=6.0cm,width=0.40\textwidth]{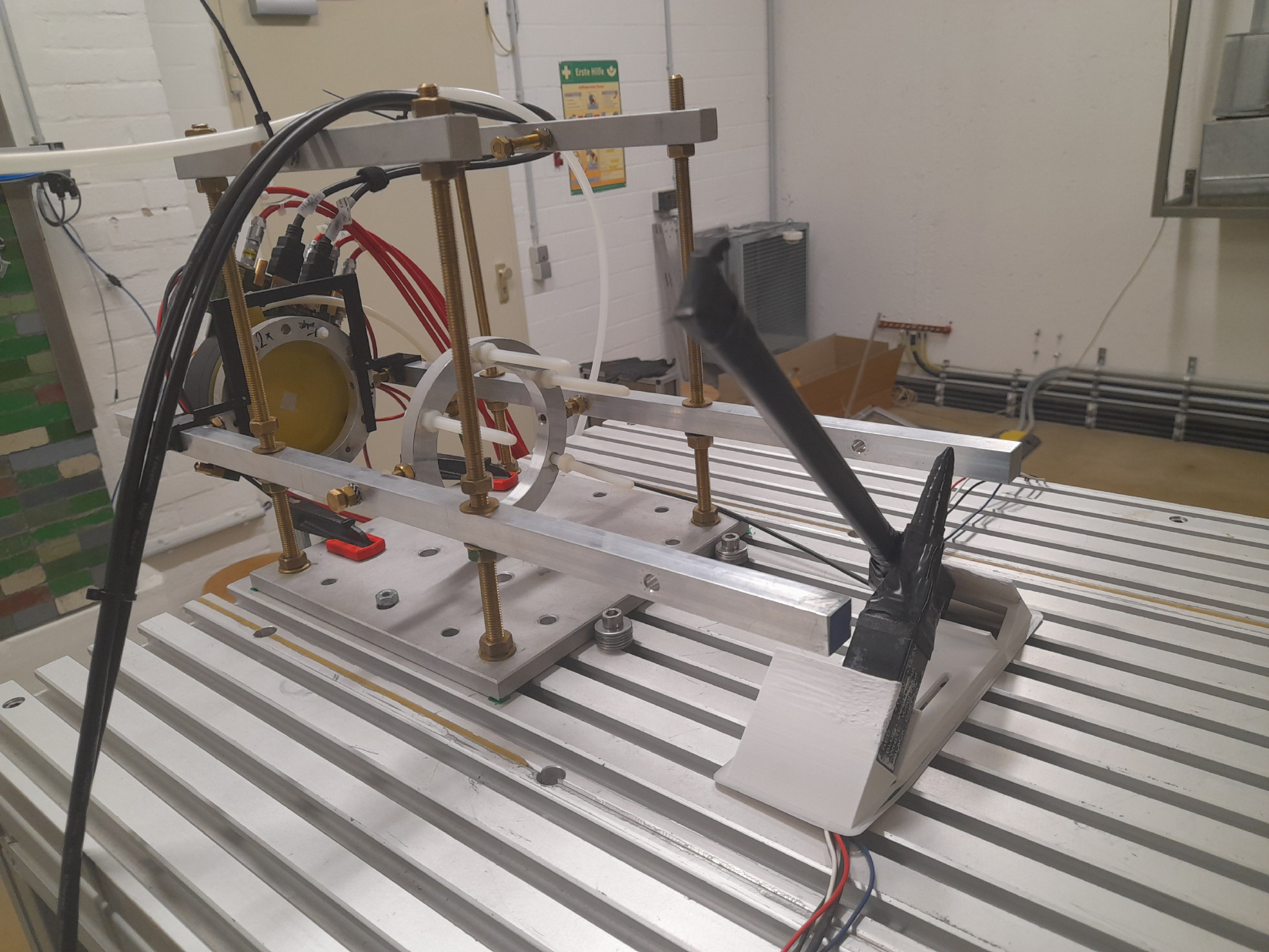}
    \includegraphics[height=6.0cm,width=0.37\textwidth]{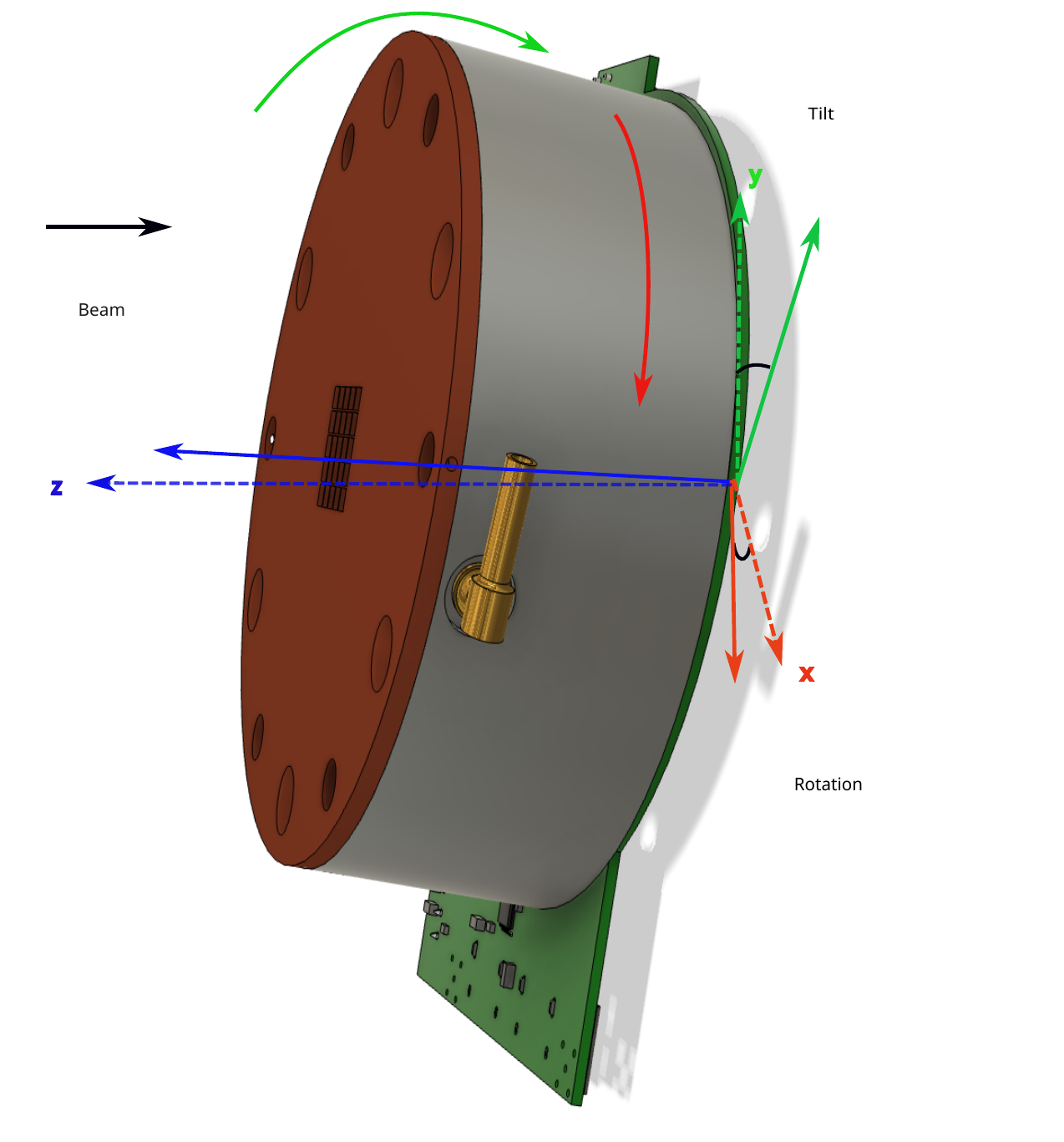}
    \includegraphics[height=6.0cm,width=0.20\textwidth]{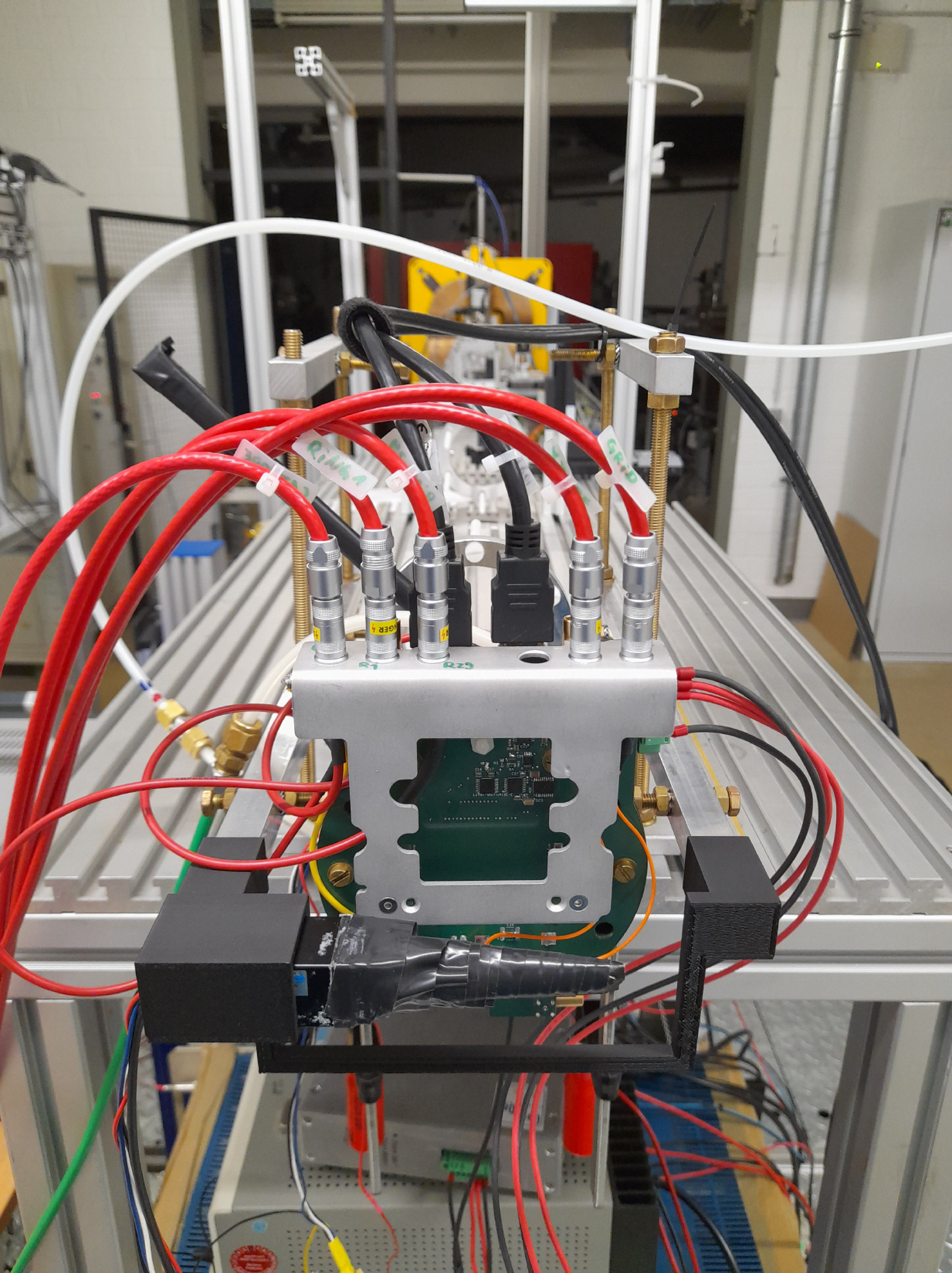}
\caption{\footnotesize Test beam setup at ELSA. Left: side view of the mechanical support used
to hold, tilt, and rotate the GridPix detector. Middle: CAD model of the GridPix detector; the green and red curved arrows indicate, respectively, the tilt and rotation directions used to obtain tracklets in different angular configurations. The coordinate systems attached to the Timepix chip before and after the application of the tilt and rotation are shown by solid and dashed lines, respectively. The dashed blue line passing through the centre of the Timepix chip defines the $z$ axis. Right: rear view of the detector hardware during a separate test beam campaign, showing the GridPix IMB, HDMI readout connections, and SHV
high-voltage connections.}
\label{fig:TB1_ELSA_setup_GridPix}
\end{figure}

To study the detector response for different tracklet directions, the GridPix
detector was mounted on a mechanical test stand designed to allow
tilt and rotation. A detector coordinate system is defined with its origin at the
centre of the Timepix ASIC, shown in Fig.~\ref{fig:TB1_ELSA_setup_GridPix} (middle). When viewed along the incident beam direction, the
positive $x$ axis points to the right and the positive $y$ axis points vertically
upwards. The $z$ axis is normal to the chip plane and defines the longitudinal
drift coordinate, with $z=0$ at the Timepix anode and positive $z$ increasing
towards the cathode and detector window.

The detector was mounted on an aluminium support structure incorporating a
rotation ring around the detector window. The tilt was set by placing dedicated
3D-printed spacers between the support frame and the rotation ring, fixing the
detector at the required inclination. In both test beams, increasing the
nominal tilt corresponded to a backward tilt. Tilting changed the beam direction relative to the detector axes, primarily in the $yz$ projection before rotation. The detector was rotated about its central $z$ axis. When viewed
from the front of the detector, clockwise rotations were defined as
positive. The desired angle was set by aligning the corresponding holes in the
detector body and rotation ring before securing the assembly. The setup for the $21\,\mathrm{mm}$ detector variant allowed
nominal tilt angles between $0^{\circ}$ and $90^{\circ}$ and rotations in
$30^{\circ}$ steps up to $60^{\circ}$. During the test beam measurements with the $5\,\mathrm{mm}$ drift-length detector, the mechanical tilt and rotation structure was modified, allowing rotations in $10^{\circ}$ steps while retaining the same backward-tilt and clockwise-rotation
convention.
For the $21\,\mathrm{mm}$ drift-length detector, measurements were performed at the nominal
tilt--rotation configurations $[0^{\circ},0^{\circ}]$,
$[2^{\circ},0^{\circ}]$, $[5^{\circ},0^{\circ}]$,
$[10^{\circ},0^{\circ}]$, $[2^{\circ},30^{\circ}]$,
$[10^{\circ},30^{\circ}]$, and $[20^{\circ},30^{\circ}]$. The dataset taken at
the nominal $[2^{\circ},30^{\circ}]$ setting is not included in the response-model
calibration. Its reconstructed mean track angles are not consistent with the
nominal detector orientation or the corresponding simulated geometry, indicating a
substantial mechanical misalignment. Since the actual orientation could not be
determined reliably after the measurement, this configuration is excluded from the common
six-configuration comparison.

For the $5\,\mathrm{mm}$ drift-length detector, only the
$[10^{\circ},20^{\circ}]$ configuration could be studied because of operational
constraints. It is used as an independent validation of the response model calibrated using the $21\,\mathrm{mm}$ measurements.
\section{Clustering and Track Reconstruction}
\label{sec:Clustering_Track_Reco}
\begin{figure}[htbp]
    \centering
    \begin{subfigure}{0.32\textwidth}
        \centering
        \includegraphics[width=\linewidth,height=4cm]{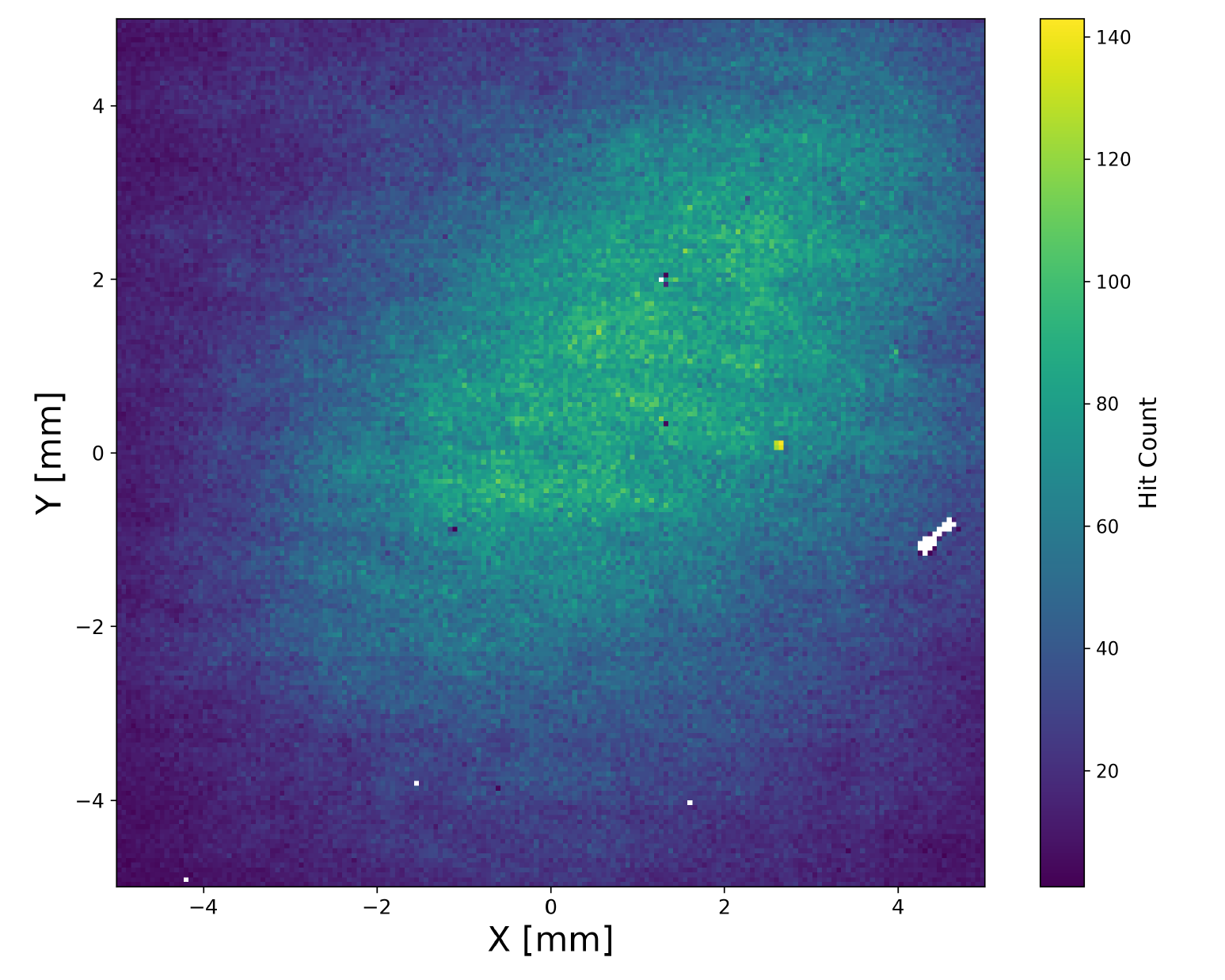}
    \end{subfigure}
    \hfill
    \begin{subfigure}{0.32\textwidth}
        \centering
        \includegraphics[width=\linewidth,height=4cm]{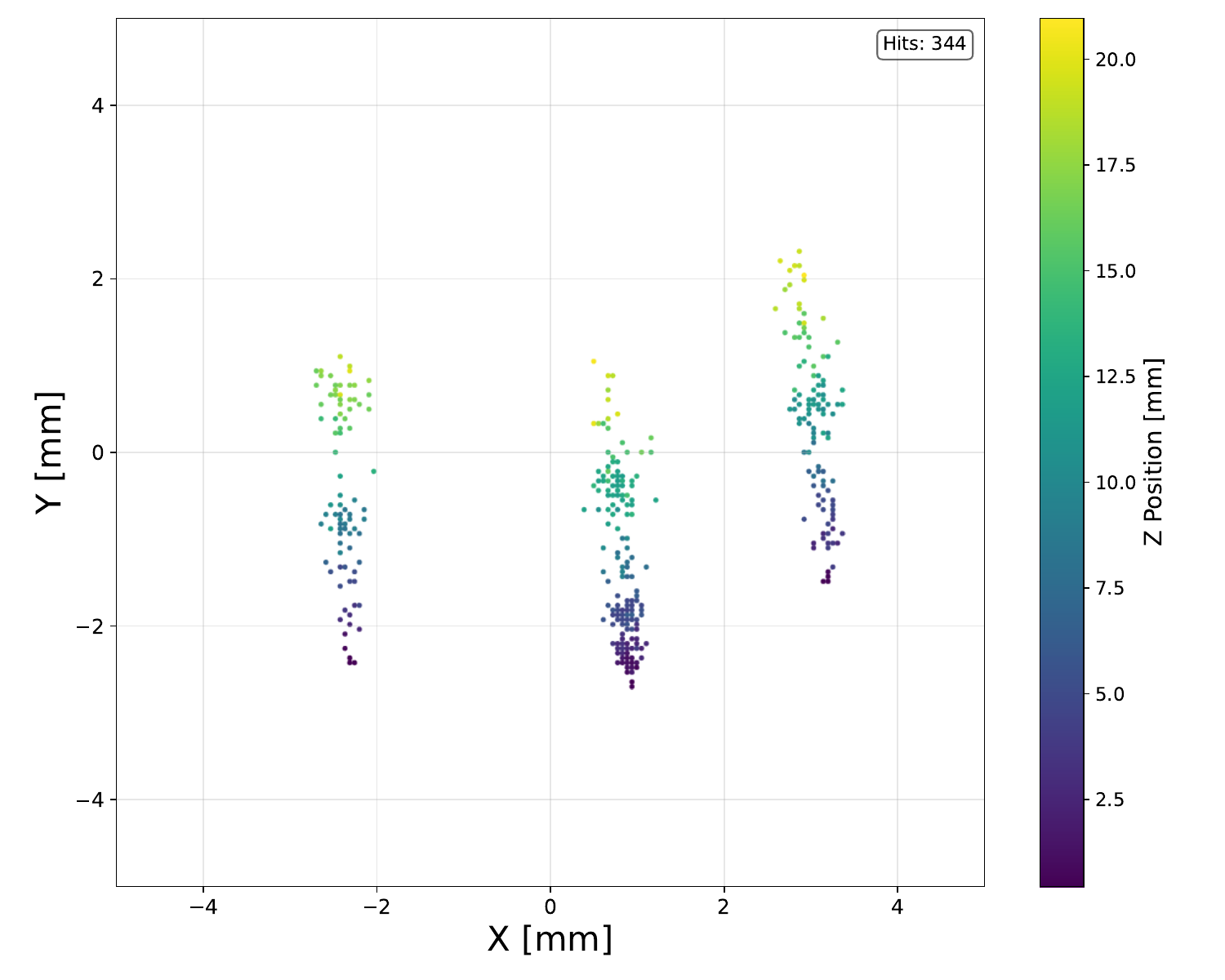}
    \end{subfigure}
    \hfill
    \begin{subfigure}{0.32\textwidth}
        \centering
        \includegraphics[width=\linewidth,height=4cm]{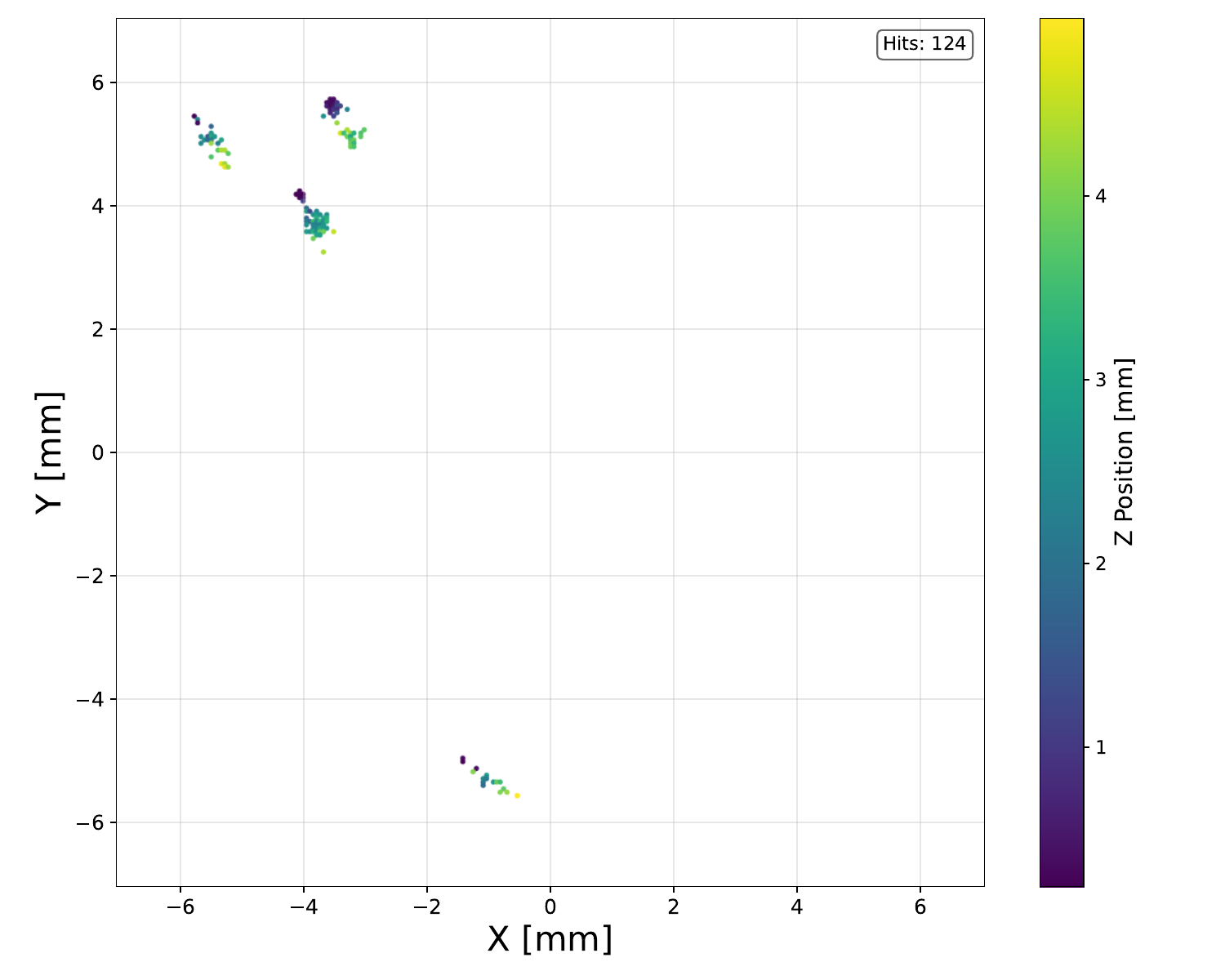}
    \end{subfigure}
    \caption{\footnotesize{Left: Pixel hit map for the $21\,\mathrm{mm}$ drift-volume detector
at $10^{\circ}$ tilt and $0^{\circ}$ rotation, for the fiducial region of $10\times10\,\mathrm{mm}^2$. Middle: Pixel hits and their reconstructed $z$ positions
for a representative event of the same configuration, with the colour scale indicating the reconstructed hit positions from the cathode (yellow) to the anode (blue). Right: Pixel hits and their reconstructed $z$ positions for a representative event of the
$5\,\mathrm{mm}$ drift-volume detector at $10^{\circ}$ tilt and $20^{\circ}$ rotation, for the full chip region, with the colour scale indicating the reconstructed hit positions from the cathode (yellow) to the anode (blue).}}
    \label{fig:occupancy}
\end{figure}

 As the detector was operated in Time-of-Arrival mode, the three-dimensional
coordinates of each hit are calculated from the recorded event information, namely the pixel hit indices and clock-cycle values. The pixel indices are converted into transverse $x$ and $y$ positions using the $55\,\mu\mathrm{m}$ pixel pitch and the chip geometry. The longitudinal coordinate, $z$, is obtained
from the electron drift time and drift velocity. The drift time is derived from the ToA value after correcting for the
shutter-window duration and trigger delay. The drift velocity is obtained from Magboltz~\cite{magboltz} simulations for the Ar:CO$_2$ (70:30) gas mixture, the applied drift field of $500\,\mathrm{V\,cm^{-1}}$, and the operating conditions during the test beam measurements. Reconstructed hits with unphysical drift times or positions outside the physical drift volume are discarded. The resulting cleaned three-dimensional hit coordinates form the input to the clustering and track-reconstruction procedure. Representative hit patterns are shown in Fig.~\ref{fig:occupancy}. The first
two panels illustrate the reconstructed hit population and a representative event for the $21\,\mathrm{mm}$ detector variant, while the third shows an event recorded with the $5\,\mathrm{mm}$ detector variant.

After translating the raw data into 3D detector coordinates, hits are clustered
within each event using a hierarchical density-based clustering algorithm
(HDBSCAN)~\cite{hdbscan}. A relaxed clustering is first applied to identify candidate track segments, followed by an
iterative merging procedure that combines nearby segments based on collinearity,
3D proximity, and $z$-continuity. For events containing multiple track candidates,
a stricter clustering pass is subsequently applied to separate close or overlapping
tracks. This produces well-defined hit clusters associated with candidate
beam-particle trajectories. 

Following clustering, hits associated with delta-electrons are rejected using residual-based
criteria, and clusters failing quality requirements on hit count, $z$-span, hit
density, and linearity are discarded. Surviving clusters are first characterised with a principal component analysis
(PCA)~\cite{PCA}, which provides an initial track direction for the subsequent Kalman filter track fit. 

\begin{figure}[htbp]
    \centering
    \begin{subfigure}{0.37\textwidth}
        \centering
        \includegraphics[width=\linewidth,height=5.9cm]{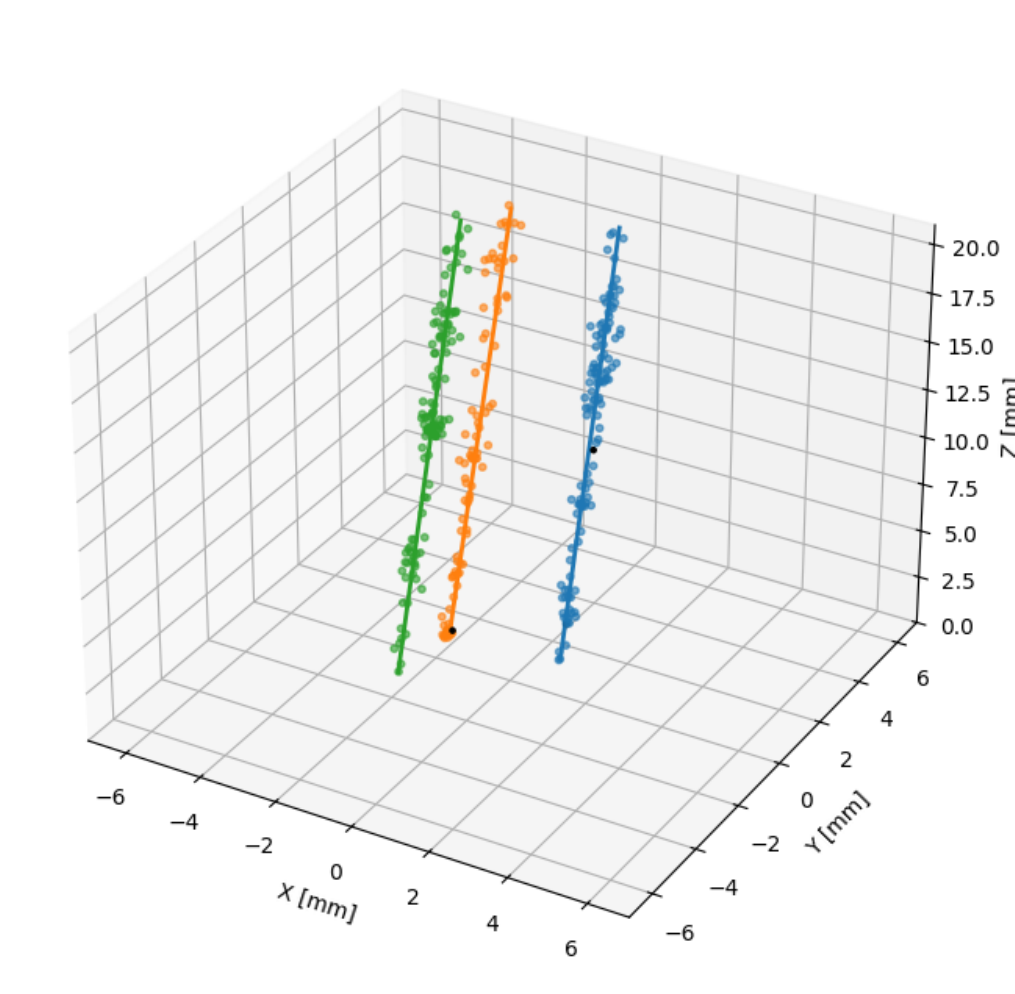}
    \end{subfigure}
    \hfill
    \begin{subfigure}{0.62\textwidth}
        \centering
        \includegraphics[width=\linewidth,height=5cm]{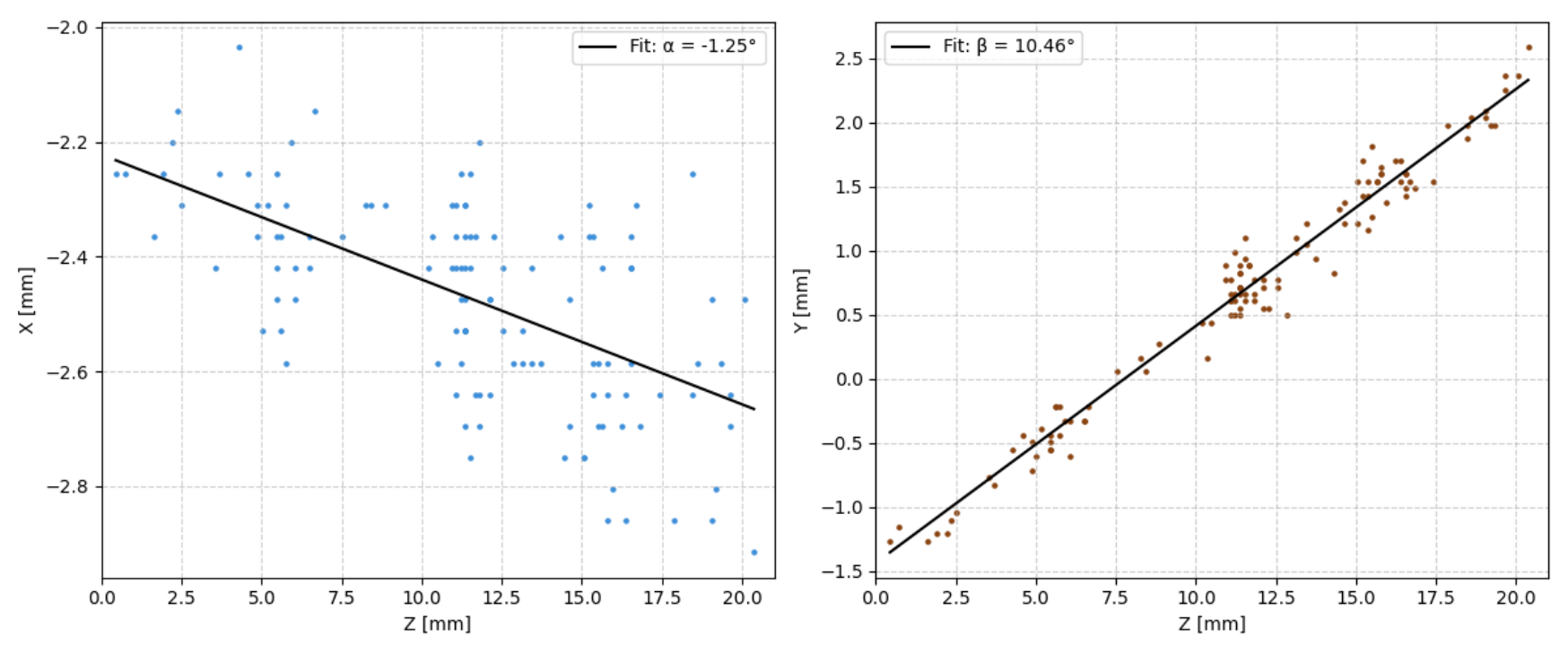}
    \end{subfigure}
    \caption{\footnotesize {Left: Fitted tracks shown for a representative event containing multiple tracks recorded in the $[10^\circ,0^\circ]$ configuration with the $21\,\mathrm{mm}$ detector. Right: $xz$ and $yz$ projections for one of the fitted tracks shown in black.}}
    \label{fig:3d_fitted_tracks_tb1}
\end{figure}

\begin{figure}[htbp]
    \centering
    \begin{subfigure}{0.37\textwidth}
        \centering
        \includegraphics[width=\linewidth,height=5.9cm]{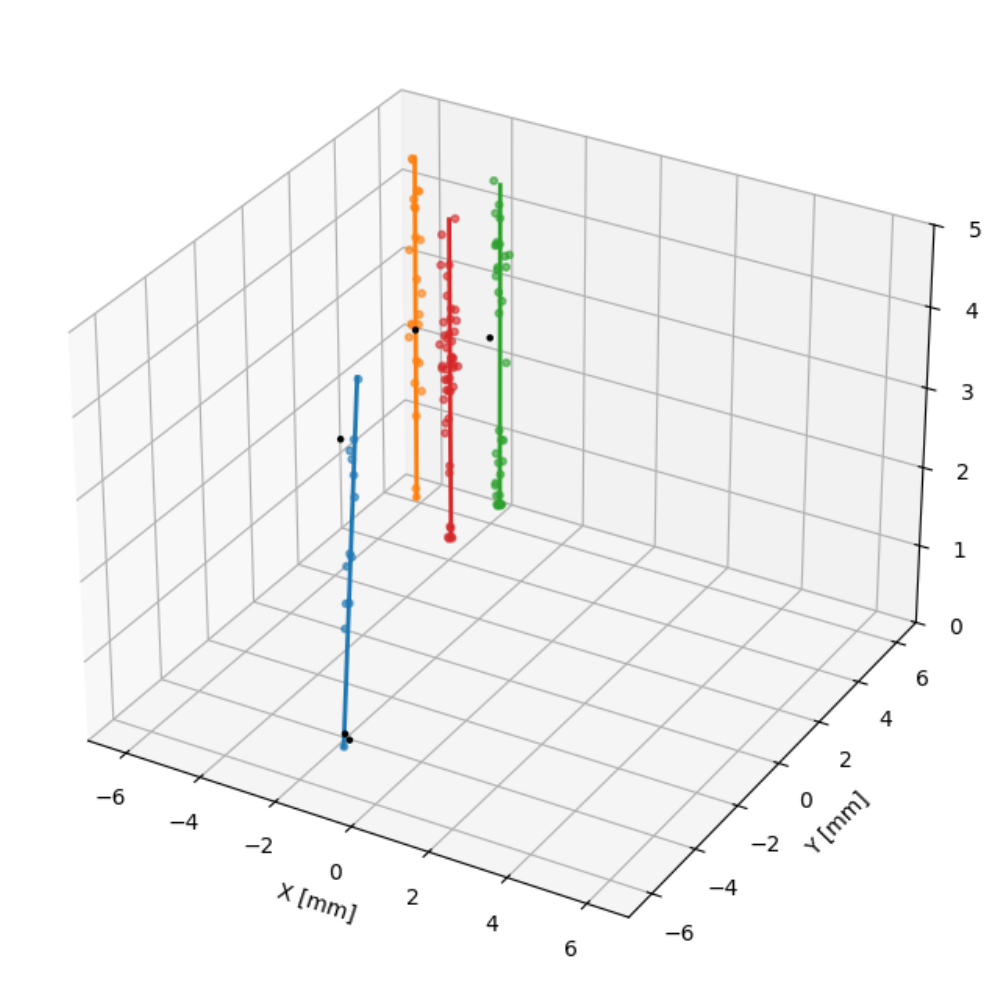}
    \end{subfigure}
    \hfill
    \begin{subfigure}{0.62\textwidth}
        \centering
        \includegraphics[width=\linewidth,height=5cm]{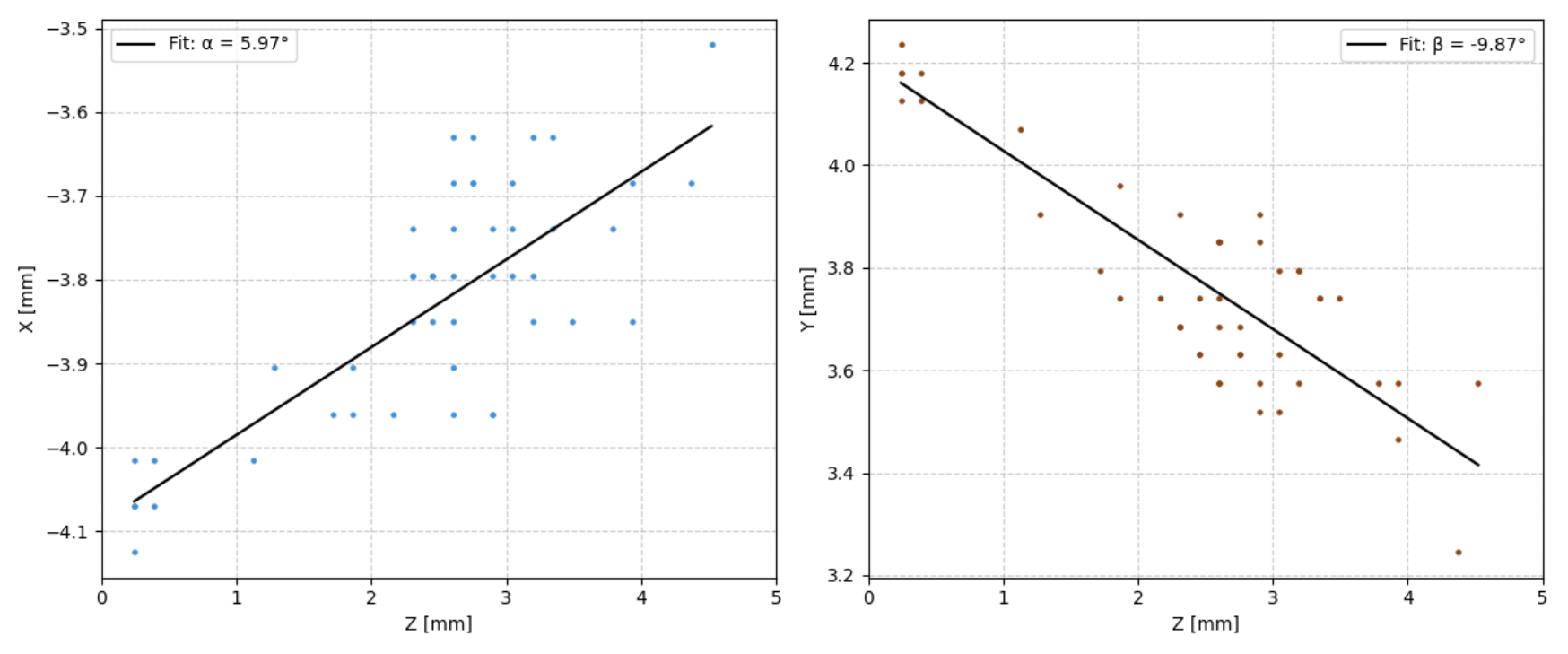}
    \end{subfigure}
    \caption{\footnotesize {Left: Fitted tracks shown for a representative event containing multiple tracks recorded in the $[10^\circ,20^\circ]$ configuration with the $5\,\mathrm{mm}$ detector. Right: $xz$ and $yz$ projections for one of the fitted tracks shown in black.}}
    \label{fig:3d_fitted_tracks_tb2}
\end{figure}

A Kalman filter is then applied to each cluster, with the fit propagated along the drift direction, $z$. The measurement covariance explicitly accounts for transverse and
longitudinal diffusion, pixel geometry, and timing uncertainty from the ToA
readout, while multiple scattering is included in the process-noise term. The quality of the Kalman filter fit is assessed using $\chi^{2}/\mathrm{ndf}$, and only tracks with $\chi^{2}/\mathrm{ndf} < 5$ are accepted. For all studied configurations, the resulting $\chi^{2}/\mathrm{ndf}$ distributions peak below 2, with peak positions around 1--1.3 for tilt angles up to $10^{\circ}$. The numerical requirements applied at the different reconstruction and selection stages are summarised in Table~\ref{tab:track_reconstruction_cuts}. Representative events containing multiple reconstructed track candidates are shown in Fig.~\ref{fig:3d_fitted_tracks_tb1} and Fig.~\ref{fig:3d_fitted_tracks_tb2} for the $21\,\mathrm{mm}$ and $5\,\mathrm{mm}$ detectors, respectively. The three-dimensional views illustrate the separation and fitting of distinct track clusters within the same event, while the $xz$ and $yz$ projections show the fitted trajectory of one representative accepted tracklet.
\begin{table*}[tb]
\centering
\small
\caption{Numerical reconstruction and selection criteria applied to the $21\,\mathrm{mm}$
data in the central $10\times10\,\mathrm{mm}^2$ fiducial region.}
\label{tab:track_reconstruction_cuts}
\renewcommand{\arraystretch}{1.15}
\begin{tabular}{lll}
\hline
\textbf{Stage} & \textbf{Requirement} & \textbf{Criterion} \\
\hline
Fiducial selection
& Central detector region
& $|x|<5\,\mathrm{mm}$, $|y|<5\,\mathrm{mm}$ \\
Merge-candidate selection
& Minimum segment size
& $N_{\mathrm{hits}}\geq 5$ \\
Final cluster selection
& Minimum number of hits
& $N_{\mathrm{hits}}\geq 10$ \\
Final cluster selection
& Longitudinal extent
& $\Delta z \geq 5\,\mathrm{mm}$ \\
Final cluster selection
& Longitudinal hit density
& $N_{\mathrm{hits}}/\Delta z \geq 0.1\,\mathrm{mm}^{-1}$ \\
Final cluster selection
& PCA linearity
& $\lambda_{\max}/\lambda_{\mathrm{mid}} > 20$ \\
Delta-electron rejection
& Normalised transverse radial residual
& $r_{\perp}/\sigma_{\perp}\leq 4$ \\
Kalman-fit selection
& Fit quality
& $\chi^{2}/\mathrm{ndf}<5$ \\
Post-fit selection
& Configuration-dependent angular window
& $|\alpha-\mu_{\alpha}|<5\sigma_{\alpha}$ and
$|\beta-\mu_{\beta}|<5\sigma_{\beta}$ \\
\hline
\end{tabular}
\vspace{1mm}
\begin{minipage}{0.95\textwidth}
\footnotesize
The PCA linearity requirement is based on the ratio of the largest to the
second-largest eigenvalue of the three-dimensional hit-position covariance
matrix; a large ratio selects elongated, approximately linear clusters.
For delta-electron rejection, transverse residuals with respect to a preliminary
PCA track are normalised by their expected hit-position uncertainties, and hits
above four standard deviations are removed. The configuration-dependent angular
window is a broad outlier-rejection requirement, where
$\mu_{\alpha,\beta}$ and $\sigma_{\alpha,\beta}$ are the mean and
standard deviation of each reconstructed angle distribution for the corresponding run. These quantities are determined before applying the corresponding angular outlier-rejection requirement.
\end{minipage}
\end{table*}

For each accepted track, the reconstructed $xz$ and $yz$ angles are defined as
\[
\alpha=\arctan\left(\frac{\mathrm{d}x}{\mathrm{d}z}\right), \qquad
\beta=\arctan\left(\frac{\mathrm{d}y}{\mathrm{d}z}\right),
\]
respectively. These are the angles of the fitted track relative to the detector $z$ axis in the $xz$ and $yz$ projections; the coordinate axes themselves remain orthogonal. The standard deviations of their distributions are used as the angular-width observables. Spatial residuals are evaluated after the Kalman fit using all retained hits included in the fitted track. For each hit, $r_x$ and $r_y$ are the transverse $x$ and $y$ components of the shortest vector from the hit to the three-dimensional fitted track line. The widths in Table~\ref{tab:full_reduced_chip_residual_widths} are standard deviations of the binned residual distributions, evaluated from bin centres weighted by bin contents. All finite residuals within the common range of $\pm500\,\mu\mathrm{m}$ enter the distributions with equal weight. Since the same hits contribute to both the residual distributions and the fitted tracks, these widths describe internal hit--track consistency and do not provide an independent measurement of the intrinsic spatial resolution.

To reduce edge effects from electric field distortions and beam misalignment, the analysis is restricted to the central $10\times10\,\mathrm{mm}^2$ fiducial region. The same fiducial region is used in the subsequent data-to-simulation optimisation studies. The angular widths and spatial-residual widths are presented in Table~\ref{tab:full_reduced_chip_angular_resolutions} and Table~\ref{tab:full_reduced_chip_residual_widths}, with the corresponding full-chip widths also given to
show the effect of the fiducial selection. The stated tilt and rotation angles refer to the nominal mechanical settings, since the absolute beam--detector orientation was not measured independently
for each run.

The angular widths are approximately $5$--$10\,\mathrm{mrad}$ for the six $21\,\mathrm{mm}$ configurations. Increasing inclination extends the track projection across the chip, but the measured widths do not improve monotonically. The central fiducial selection mainly improves $\sigma_\beta$, while $\sigma_\alpha$ changes little for several unrotated runs. A beam offset along $y$, combined with track elongation in that direction, can cause cathode-side hits to fall outside the chip. The resulting shorter longitudinal lever arm makes the $yz$ slope more sensitive to diffusion, timing, and hit-population fluctuations. Fully contained tracklets retain a more uniform lever arm; the smaller truncation along $x$ explains the weaker change in $\alpha$. Fiducial-area sensitivity enters the data uncertainty in Sec.~\ref{sec:uncertainty_estimation}.

\begin{table*}[tb]
\centering

\begin{minipage}[t]{0.49\textwidth}
\vspace{0pt}
\centering
\caption{Angular widths for the full chip and central
$10\times10\,\mathrm{mm}^2$ fiducial region.}
\label{tab:full_reduced_chip_angular_resolutions}
\footnotesize
\setlength{\tabcolsep}{3pt}
\renewcommand{\arraystretch}{1.25}
\resizebox{\linewidth}{!}{%
\begin{tabular}{rrrrrr}
\hline
\shortstack{Drift\\length} &
\shortstack{Tilt,\\rotation} &
$\sigma_{\alpha}^{\mathrm{full}}$ &
$\sigma_{\beta}^{\mathrm{full}}$ &
$\sigma_{\alpha}^{\mathrm{reduced}}$ &
$\sigma_{\beta}^{\mathrm{reduced}}$ \\
$[\mathrm{mm}]$ & $[^\circ]$ &
$[\mathrm{mrad}]$ & $[\mathrm{mrad}]$ &
$[\mathrm{mrad}]$ & $[\mathrm{mrad}]$ \\
\hline
21 & 0,0   & 5.4  & 5.5  & 5.0  & 4.9  \\
21 & 2,0   & 4.4  & 8.4  & 4.5  & 5.4  \\
21 & 5,0   & 4.4  & 9.6  & 4.5  & 5.5  \\
21 & 10,0  & 5.6  & 7.6  & 5.5  & 5.8  \\
21 & 10,30 & 5.3  & 9.4  & 5.3  & 6.8  \\
21 & 20,30 & 7.6  & 10.7 & 7.1  & 9.9  \\
5  & 10,20 & 22.9 & 30.0 & 22.5 & 29.3 \\
\hline
\end{tabular}%
}
\end{minipage}
\hfill
\begin{minipage}[t]{0.49\textwidth}
\vspace{0pt}
\centering
\caption{Spatial-residual widths for the full chip and central
$10\times10\,\mathrm{mm}^2$ fiducial region.}
\label{tab:full_reduced_chip_residual_widths}
\footnotesize
\setlength{\tabcolsep}{3pt}
\renewcommand{\arraystretch}{1.15}
\resizebox{\linewidth}{!}{%
\begin{tabular}{rrrrrr}
\hline
\shortstack{Drift\\length} &
\shortstack{Tilt,\\rotation} &
$\sigma_{r_x}^{\mathrm{full}}$ &
$\sigma_{r_y}^{\mathrm{full}}$ &
$\sigma_{r_x}^{\mathrm{reduced}}$ &
$\sigma_{r_y}^{\mathrm{reduced}}$ \\
$[\mathrm{mm}]$ & $[^\circ]$ &
$[\mu\mathrm{m}]$ & $[\mu\mathrm{m}]$ &
$[\mu\mathrm{m}]$ & $[\mu\mathrm{m}]$ \\
\hline
21 & 0,0   & 155 & 160 & 153 & 151 \\
21 & 2,0   & 140 & 143 & 137 & 133 \\
21 & 5,0   & 139 & 149 & 136 & 143 \\
21 & 10,0  & 143 & 161 & 140 & 158 \\
21 & 10,30 & 147 & 157 & 145 & 155 \\
21 & 20,30 & 151 & 182 & 149 & 181 \\
5  & 10,20 & 83  & 93  & 81  & 91  \\
\hline
\end{tabular}%
}
\end{minipage}

\end{table*}

\FloatBarrier

\section{Monte Carlo Simulation of the GridPix Detector}
\label{sec:MC_Geant4_Simulation}
 Alongside the track-reconstruction algorithm, a Geant4-based simulation was developed to construct a response model that translates the truth information available from the Monte Carlo simulation into a realistic detector output. The combination of the Geant4 simulation and the detector-response model can then be used to characterise the behaviour of the GridPix detector for different angular configurations.
 
The simulation replicates the cylindrical detector geometry employed in the test beam campaigns, with a diameter of $78\,\mathrm{mm}$ and a $21\,\mathrm{mm}$ drift length, filled with an Ar:CO$_2$ mixture in a 70:30 ratio. The sensitive gas volume is divided into an array of columns extending over the full drift length, each with a square cross-sectional area of $55\times55\,\mu\mathrm{m}^2$, corresponding to the Timepix pixel dimensions. The simulated accelerator electrons are monoenergetic, with an energy of $2.6\,\mathrm{GeV}$. For the independent $5\,\mathrm{mm}$ comparison, the corresponding detector
geometry was used with the drift length reduced to $5\,\mathrm{mm}$.

During the measurements, the ELSA electron beam was reported to have angular
divergences of $\sigma_{x,\mathrm{beam}}=0.5\,\mathrm{mrad}$ and
$\sigma_{y,\mathrm{beam}}=0.2\,\mathrm{mrad}$ in the two directions transverse
to the nominal beam axis. These values are approximate beam-divergence estimates communicated by the
ELSA beamline team when the beam properties were requested for this study.
They were not obtained from dedicated measurements, and no independent
uncertainties were provided. The provided beam divergences are introduced into the simulation on an event-by-event basis during the generation of beam electrons, as independent Gaussian perturbations to the primary electron direction in the transverse directions,
\begin{equation}
\begin{gathered}
\delta\theta_{x,\mathrm{beam}} \sim
\mathcal{N}(0,\sigma_{x,\mathrm{beam}}),
\qquad
\delta\theta_{y,\mathrm{beam}} \sim
\mathcal{N}(0,\sigma_{y,\mathrm{beam}}).
\end{gathered}
\end{equation}
The two perturbations contribute primarily to the simulated $xz$ and $yz$
track projections, respectively. Because the divergence is sampled
independently for each primary electron at the point of event generation, its
effect propagates through the full simulation chain and is already present in
the simulated ionisation-electron distributions used in the subsequent
analysis.

The simulation employs the standard \texttt{G4EmStandardPhysics\_option4} electromagnetic physics constructor, supplemented by the \texttt{G4PAIPhotModel}~\cite{g4pai}. The latter implements the Photo-Absorption-ionisation (PAI) model and replaces the default ionisation treatment for charged particles traversing the low-density gaseous medium. The PAI model calculates energy deposition and $\delta$-ray
production from the photoabsorption cross section of the gas, providing a
more detailed description of ionisation-energy-loss fluctuations and
secondary-electron production in the gaseous medium. For this simulation study, a production range cut of $0.02\,\mathrm{mm}$ was applied. The density of the gas mixture was set to $1.712\,\mathrm{mg/cm^3}$.

The gas detector geometry is positioned within the Geant4 world volume such
that the chip centre coincides with the origin of the global coordinate
system. The pixel columns are constructed as daughter volumes within the gas
detector geometry, so that any transformation applied to the gas volume is
inherited automatically by the entire pixel array. Detector tilt and rotation
are implemented as transformations of the gas detector geometry itself, with
the axes of both transformations passing through the chip centre. The two
transformations are applied sequentially: the detector is first tilted
backwards, following the convention adopted in the test beam analysis, and
subsequently rotated. Since the Monte Carlo truth information is recorded by
Geant4 in the global coordinate system, the ionisation positions are first
transformed into the local coordinate system of the gas volume by inverting
the placement rotation and translation. Because this local frame is centred
on the geometric mid-plane of the drift gap rather than on the chip, the
$z$ coordinate is subsequently shifted by half the drift-gap height such that
the origin coincides with the chip plane and $z$ increases towards the
cathode. The resulting simulation truth dataset consists of the
three-dimensional Monte Carlo truth positions of the recorded ionisation
sites along the trajectory of the incident beam electron.
\section{Detector-Response Model and Global Optimisation}
\label{sec:Detector_Response_Model}
\subsection{Known detector response}
\label{sec:known_detector_response}
 The three-dimensional truth information from the simulation does not directly correspond to observed GridPix hits. The truth ionisation positions follow the trajectories of the beam electron and the produced secondary particles, but do not include electron transport in the gas or detector-response effects. A detector-response model is therefore applied event by event to translate the Geant4 truth information into reconstructed detector coordinates.
 
The first response component introduces transverse and longitudinal diffusion.
For a drift distance $z$, the corresponding widths are
\begin{equation}
\sigma_{\mathrm{T}}(z) =
\sqrt{2D_{\mathrm{T}}\frac{z}{v_{\mathrm{d}}}},
\qquad
\sigma_{\mathrm{L}}(z) =
\sqrt{2D_{\mathrm{L}}\frac{z}{v_{\mathrm{d}}}},
\label{eq:diffusion_response}
\end{equation}
where $D_{\mathrm{T}}=105.97\,\mathrm{cm^2\,s^{-1}}$ and $D_{\mathrm{L}}=117.4\,\mathrm{cm^2\,s^{-1}}$ are the transverse and longitudinal diffusion coefficients derived from the Magboltz-predicted RMS diffusion spreads for the Ar:CO$_2$ (70:30) gas mixture at the applied drift field of $500\,\mathrm{V\,cm^{-1}}$, and $v_{\mathrm{d}}=(1.181\pm0.003)\,\mathrm{cm\,\mu s^{-1}}$ is the corresponding electron drift velocity.
The diffusion-only widths in Eq.~\ref{eq:diffusion_response} can equivalently be written as $\sigma_{\mathrm{T,L}}(z)=C_{\mathrm{T,L}}\sqrt{z}$, with $C_{\mathrm{T,L}}=\sqrt{2D_{\mathrm{T,L}}/v_{\mathrm{d}}}$. The values above correspond to $C_{\mathrm{T}}=(134\pm7)\,\mu\mathrm{m}/\sqrt{\mathrm{cm}}$ and $C_{\mathrm{L}}=(141\pm6)\,\mu\mathrm{m}/\sqrt{\mathrm{cm}}$. A total hit-width parametrisation may also contain a drift-independent contribution, $\sigma_{\mathrm{tot},j}^2(z)=\sigma_{\mathrm{const},j}^2+C_j^2z$, where $j=\mathrm{T,L}$. Here Eq.~\ref{eq:diffusion_response} specifies the diffusion contribution; pixelisation, timing and the subsequent empirical smearing are described separately.

The uncertainties on the nominal diffusion coefficients can be obtained
from the uncertainties on the corresponding diffusion constants
$C_{\mathrm{T,L}}$ and on the electron drift velocity. From
\begin{equation}
    C_j = \sqrt{\frac{2D_j}{v_{\mathrm{d}}}},
    \qquad
    D_j = \frac{C_j^2 v_{\mathrm{d}}}{2},
    \qquad j=\mathrm{T,L},
\end{equation}
and assuming that the uncertainties on $C_j$ and $v_{\mathrm{d}}$ are
independent, standard error propagation gives
\begin{equation}
    \left(\frac{\delta D_j}{D_j}\right)^2
    =
    \left(2\frac{\delta C_j}{C_j}\right)^2
    +
    \left(\frac{\delta v_{\mathrm{d}}}{v_{\mathrm{d}}}\right)^2 .
    \label{eq:diffusion_uncertainty}
\end{equation}
For $v_{\mathrm{d}}=(1.181\pm0.003)\,\mathrm{cm\,\mu s^{-1}}$, the
relative uncertainty on the drift velocity is approximately $0.25\%$.
Using $C_{\mathrm{T}}=(134\pm7)\,\mu\mathrm{m}/\sqrt{\mathrm{cm}}$
and $C_{\mathrm{L}}=(141\pm6)\,\mu\mathrm{m}/\sqrt{\mathrm{cm}}$, the
relative uncertainties on the nominal diffusion coefficients are
\begin{equation}
    \frac{\delta D_{\mathrm{T}}}{D_{\mathrm{T}}}
    \simeq 10.5\%,
    \qquad
    \frac{\delta D_{\mathrm{L}}}{D_{\mathrm{L}}}
    \simeq 8.5\%.
\end{equation}
For the nominal Magboltz predictions, these correspond to
\begin{equation}
    D_{\mathrm{T}}^{\mathrm{nom}}
    =
    (105.97\pm11.1)\,\mathrm{cm^2\,s^{-1}},
    \qquad
    D_{\mathrm{L}}^{\mathrm{nom}}
    =
    (117.4\pm10.0)\,\mathrm{cm^2\,s^{-1}}.
    \label{eq:nominal_diffusion_uncertainties}
\end{equation}

After application of the known detector-response model, the simulated
X- and Y-residual distributions were found to be broader than the
corresponding test beam distributions for most of the $21\,\mathrm{mm}$
configurations. Since the subsequent empirical response terms introduce
additional broadening, this motivated a dedicated calibration of the
transverse diffusion coefficient before determining these additional
response parameters. The calibration was restricted to $D_{\mathrm{T}}$, since the observed
discrepancy concerns the X- and Y-residual widths, which are directly
affected by transverse diffusion. The longitudinal diffusion coefficient
$D_{\mathrm{L}}$, which primarily affects the reconstructed drift
coordinate, was therefore kept fixed at its nominal Magboltz value of
$117.4\,\mathrm{cm^2\,s^{-1}}$.

The transverse diffusion coefficient was scanned from $0.70$ to $1.30$ times its nominal value in steps of $0.02$, while $D_{\mathrm{L}}$ was kept fixed at its nominal value. The twelve X- and Y-residual widths from the six configurations were assigned equal weight. The reconstructed angular widths were retained only as diagnostics and did not contribute to the determination of $D_{\mathrm{T}}$. For each tested value of $D_{\mathrm{T}}$, the simulated X- and
Y-residual widths were compared with the corresponding test beam widths
for the six $21\,\mathrm{mm}$ configurations. The calibration minimized
\begin{equation}
    S(D_{\mathrm{T}})
    =
    \frac{1}{12}
    \sum_{r=1}^{6}
    \left(
        \frac{
        \left|\sigma_{X,r}^{\mathrm{sim}}
        -\sigma_{X,r}^{\mathrm{TB}}\right|}
        {\sigma_{X,r}^{\mathrm{TB}}}
        +
        \frac{
        \left|\sigma_{Y,r}^{\mathrm{sim}}
        -\sigma_{Y,r}^{\mathrm{TB}}\right|}
        {\sigma_{Y,r}^{\mathrm{TB}}}
    \right).
    \label{eq:DT_calibration_score}
\end{equation}

Here $r$ runs over the six configurations, and $\sigma_{X,Y}^{\mathrm{sim,TB}}$ denote simulated and measured residual widths. Each term is an absolute relative difference, with equal weight for both coordinates and all runs. This stage uses only the transverse residuals, before calibrating the additional empirical response. The minimum occurs at
\begin{equation}
    D_{\mathrm{T}}^{\mathrm{cal}}
    =93.3\,\mathrm{cm^2\,s^{-1}},
\end{equation}
corresponding to approximately $0.88$ times the nominal Magboltz value.
The central physical diffusion model used in the subsequent response
optimisation is therefore defined by
\begin{equation}
    D_{\mathrm{T}}=93.3\,\mathrm{cm^2\,s^{-1}},
    \qquad
    D_{\mathrm{L}}=117.4\,\mathrm{cm^2\,s^{-1}}.
    \label{eq:central_diffusion_model}
\end{equation}
The propagation of the uncertainties associated with these diffusion
coefficients to the final reconstructed observables, and their combination
with the other simulation uncertainties, is discussed in
Section~\ref{sec:uncertainty_estimation}.

In Time-of-Arrival mode, each Timepix pixel records a single time associated with its first threshold crossing. Consequently, when multiple simulated electrons are assigned to the same pixel within one event, only the earliest simulated arrival is retained. This behaviour, hereafter referred to as first-electron arrival (FEA), is reproduced in the simulated dataset. After transverse diffusion, several simulated electrons can be assigned to the same pixel while having different drift distances from the chip. In the present response model, the electron with the smallest drift distance is retained as the recorded hit, while subsequent electrons assigned to the same pixel are discarded.
\begin{figure}[htbp]
    \centering
    \includegraphics[trim=0 0pt 0 50pt, clip, width=0.75\linewidth]{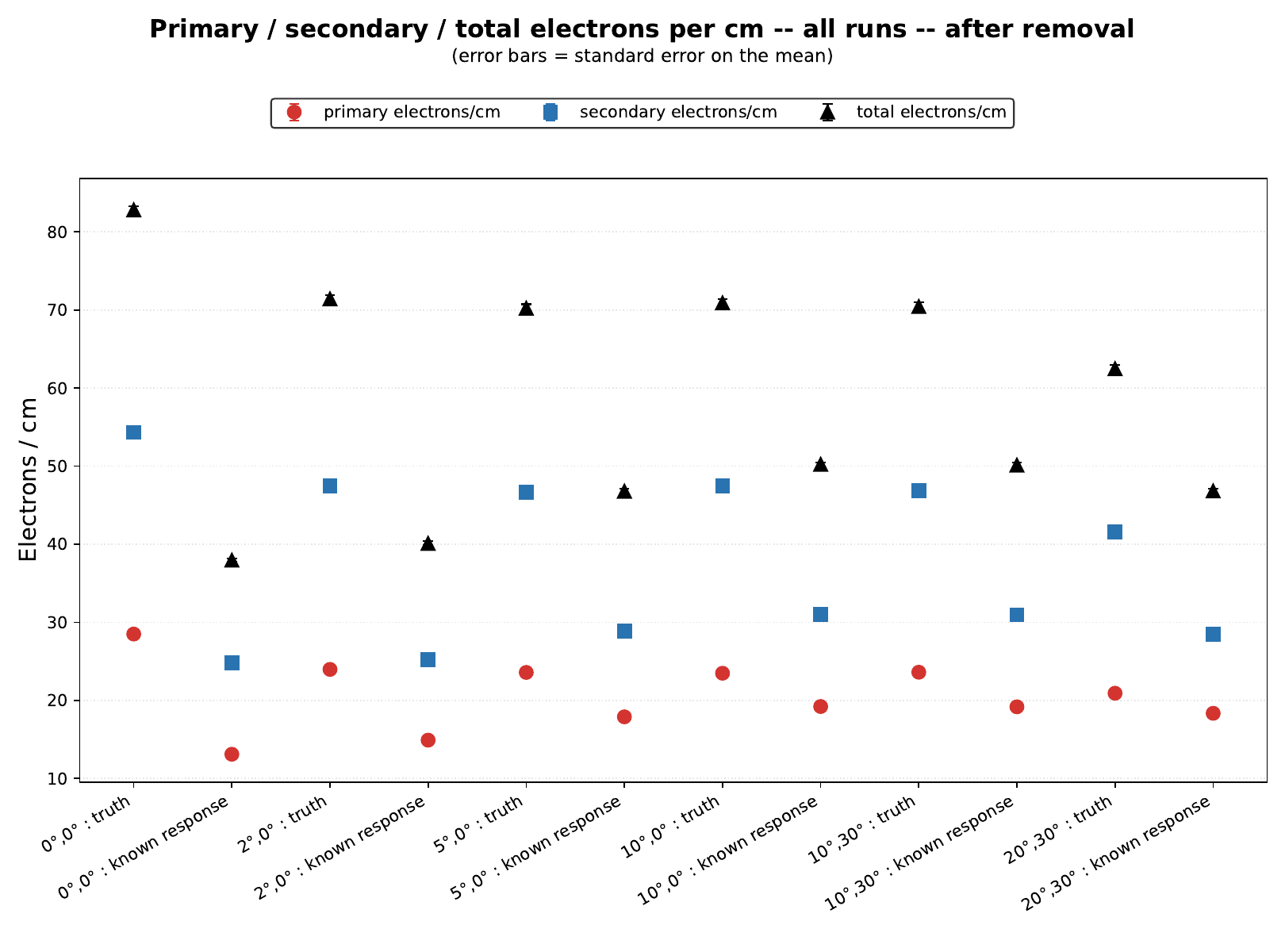}
    \caption{Primary, secondary, and total ionisation-electron yields per unit track length obtained from the truth-level simulation and after application of the known detector response for the analysed angular configurations.}
    \label{fig:electron_counts_simulation}
\end{figure}

Fig.~\ref{fig:electron_counts_simulation} shows the mean primary and secondary ionisation-electron yields per unit track length obtained from the truth-level simulation and after application of the known detector response, including FEA, for each angular configuration. At truth level, the simulation yields approximately 20--30 primary electrons per centimetre, which produce approximately 40--50 secondary electrons per centimetre. Both values are in close agreement with expectations for an Ar:CO$_2$ (70:30) gas mixture~\cite{gas_multiplicities}. Application of the known detector response reduces both populations before track reconstruction. The error bars in Fig.~\ref{fig:electron_counts_simulation} represent statistical uncertainties only.

\begin{figure}[htbp]
    \centering
    \begin{subfigure}[t]{0.48\textwidth}
        \centering
        \includegraphics[
            page=1,
            trim=0 20pt 0 30pt,
            clip,
            width=\textwidth,
            height=6.20cm
        ]{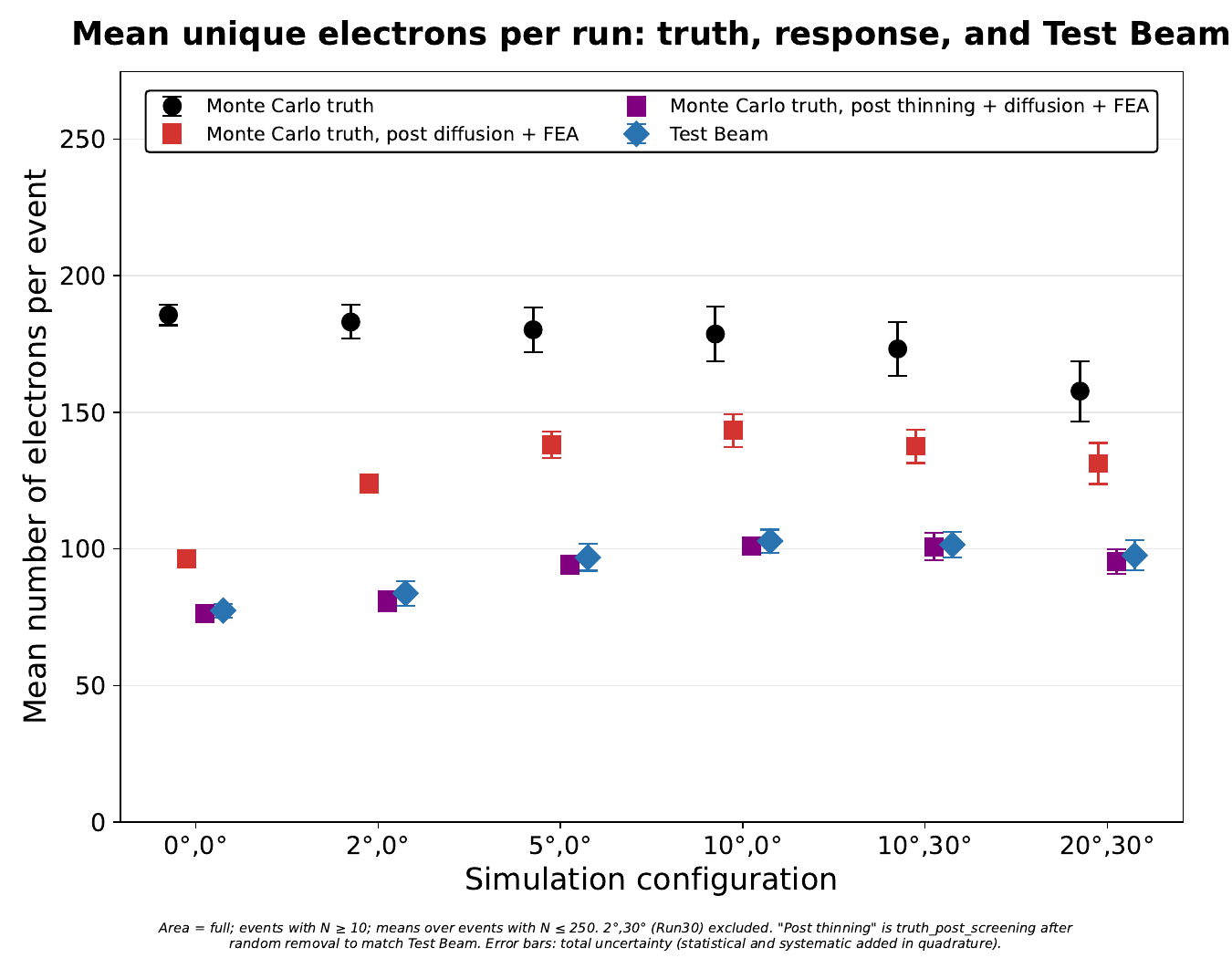}
        \caption{}
        \label{fig:tb1_sim_multiplicity_means}
    \end{subfigure}
    \hfill
    \begin{subfigure}[t]{0.48\textwidth}
        \centering
        \includegraphics[
            page=5,
            trim=0 0pt 0 22pt,
            clip,
            width=\textwidth
        ]{images/monte_carlo/Run27_10tilt0rot_electron_multiplicity_after_removal.pdf}
        \caption{}
        \label{fig:tb1_sim_multiplicity_distribution}
    \end{subfigure}
    \caption{\footnotesize
    (a) Mean electron multiplicity per event for all analysed angular
    configurations of the $21\,\mathrm{mm}$ drift-length detector: test beam
    data (blue), raw Monte Carlo truth (black), simulation after diffusion and
    FEA (red), and multiplicity-matched simulation after the additional
    thinning procedure (purple). Means are evaluated for
    $10\leq N\leq250$ according to Eq.~\ref{eq:mean_multiplicity}. Error bars
    denote the total uncertainty defined in Eq.~\ref{eq:syst}.
    (b) Electron-multiplicity distribution for the
    $[10^\circ,0^\circ]$ configuration, measured in the test beam before
    applying the lower multiplicity requirement (black, dotted) and with
    $10\leq N\leq250$ (blue), together with the multiplicity-matched
    simulation for $10\leq N\leq250$ (red). The distributions are normalised
    to unit area and use a bin width of one electron.}
    \label{fig:tb1_sim_multiplicity}
\end{figure}

Fig.~\ref{fig:tb1_sim_multiplicity}(\subref{fig:tb1_sim_multiplicity_means})
compares the mean electron multiplicity per event in the
$21\,\mathrm{mm}$ test beam data with that obtained from the simulation. Only events with at least 10 electrons on the full chip are considered, which excludes empty and low-multiplicity events observed in the test beam data as described in Fig.~\ref{fig:tb1_sim_multiplicity}(\subref{fig:tb1_sim_multiplicity_distribution}). The mean is computed over events with at most 250 electrons to limit the influence of the $\delta$-ray tail,
\begin{equation}
  \langle N \rangle = \frac{1}{|\mathcal{E}|} \sum_{e \in \mathcal{E}} N_e ,
  \qquad
  \mathcal{E} = \{\, e : 10 \le N_e \le 250 \,\}.
  \label{eq:mean_multiplicity}
\end{equation}
The selection is applied identically to the test beam data and to the simulation after the known detector response. For the raw Monte Carlo sample, the mean multiplicity is evaluated using the same simulated event identities selected after the known detector response, while the raw truth multiplicities themselves are not subjected to the multiplicity limits.

The raw Monte Carlo simulation yields 158--186 electrons per event, approximately twice the 77--103 electrons per event measured in the test beam. Application of transverse and longitudinal diffusion followed by FEA reduces this discrepancy considerably. The effect is largest for perpendicular tracks in the $[0^\circ,0^\circ]$ configuration, where most electrons are concentrated within a small number of pixels and FEA discards approximately $48\%$ of them. At larger angles, the track projection onto the chip covers a larger area, increasing the number of occupied pixels and reducing the FEA loss to approximately $17\%$ for the $[20^\circ,30^\circ]$ configuration. Nevertheless, the processed simulation multiplicities still exceed those measured in the test beam by approximately $25\%$ for the $[0^\circ,0^\circ]$ configuration and by $34$--$48\%$ for the tilted configurations.

The remaining excess indicates additional electron losses that are not included in the available detector simulation, for which electron attachment is one possible contribution. To obtain a multiplicity-matched simulation input for the subsequent response-model comparison, the simulated electrons surviving diffusion and FEA are subjected to an empirical stochastic-removal procedure. For event $e$, the electron-retention probability is
\begin{equation}
  q_e = q\left(1+r\xi_e\right),
  \qquad
  \xi_e\sim\mathcal{N}(0,1),
  \label{eq:thinning}
\end{equation}
where the resulting value of $q_e$ is restricted to the interval $[0,1]$. Here, $q$ is the mean retention probability and $r$ controls its relative event-to-event variation. Once $q_e$ has been drawn for an event, each simulated electron in that event is retained independently with probability $q_e$. For each configuration, $q$ and $r$ are determined iteratively such that the most probable value and width of the simulated multiplicity distribution, obtained from a Gaussian fit to its peak, reproduce those of the test beam distribution.
The test beam data also contain an excess of events below the multiplicity peak that is not reproduced by the simulation. Depending on the angular configuration, this excess accounts for approximately $6$--$22\%$ of the selected test beam events. An additional empirical low-multiplicity correction is therefore applied to the corresponding fraction of simulated events. This correction is determined from the difference between the test beam and simulated distributions below the multiplicity peak and is used only to construct the multiplicity-matched simulation sample.

For all tilted configurations, consistent values of $q = 0.74$--$0.77$ are
obtained, i.e.\ an electron loss of 23--26\%, so the angular dependence of the
multiplicity is dominated by diffusion and FEA. Only
$[0^{\circ},0^{\circ}]$, where FEA removes the largest fraction, requires a
higher $q = 0.81$.
The resulting multiplicity distribution is shown in
Fig.~\ref{fig:tb1_sim_multiplicity}(\subref{fig:tb1_sim_multiplicity_distribution}). After the imposed multiplicity matching, the mean simulated multiplicities agree with the corresponding test beam values within approximately $1$--$3.5\%$. This agreement quantifies the accuracy of the matching procedure and is not used as an independent validation of the detector-response model.

Each quantity described in Fig.~\ref{fig:tb1_sim_multiplicity} is assigned a statistical and a systematic uncertainty. For the
test beam, the raw Monte Carlo truth and the simulation after the known
detector response, the statistical uncertainty is the standard error of the
mean. The thinned simulation and the retention probability $q$, in contrast,
result from the matching to the test beam, and therefore also fluctuate with
the test beam sample and with the random thinning. Their statistical
uncertainty is estimated with a bootstrap: the test beam and simulated events
are resampled with replacement 100 times, the complete matching is repeated
for each resampled set, and the standard deviation of the resulting thinned
means and values of $q$ is taken.
Systematic uncertainties arise from choices made in the analysis and are
estimated by repeating it with one choice changed at a time. For all
quantities, the lower selection edge is varied between $N_{\min} = 5$ and 15
and the upper edge between $N_{\max} = 200$ and 300. For the thinned simulation
and $q$, the peak region of the Gaussian fit (0.6 and 0.8 instead of 0.7 of
the maximum) and the upper edge of the low-multiplicity correction (0.9 and 1.1
times the MPV) are varied in addition. For each source $k$, the largest deviation $\Delta_k$ from the nominal result is taken. The different contributions are treated as independent for this estimate and combined in quadrature:
\begin{equation}
  \sigma_{\mathrm{tot}} =
  \sqrt{
  \sigma_{\mathrm{stat}}^2+
  \sum_k\Delta_k^2
  }.
  \label{eq:syst}
\end{equation}
The total uncertainty amounts to 2.5--5.6 electrons for the test beam means,
2.0--5.0 electrons for the thinned simulation and 0.010--0.028 for $q$. It is
dominated by the choice of the selection edges, while the statistical
contribution is small.

\begin{figure}[htbp]
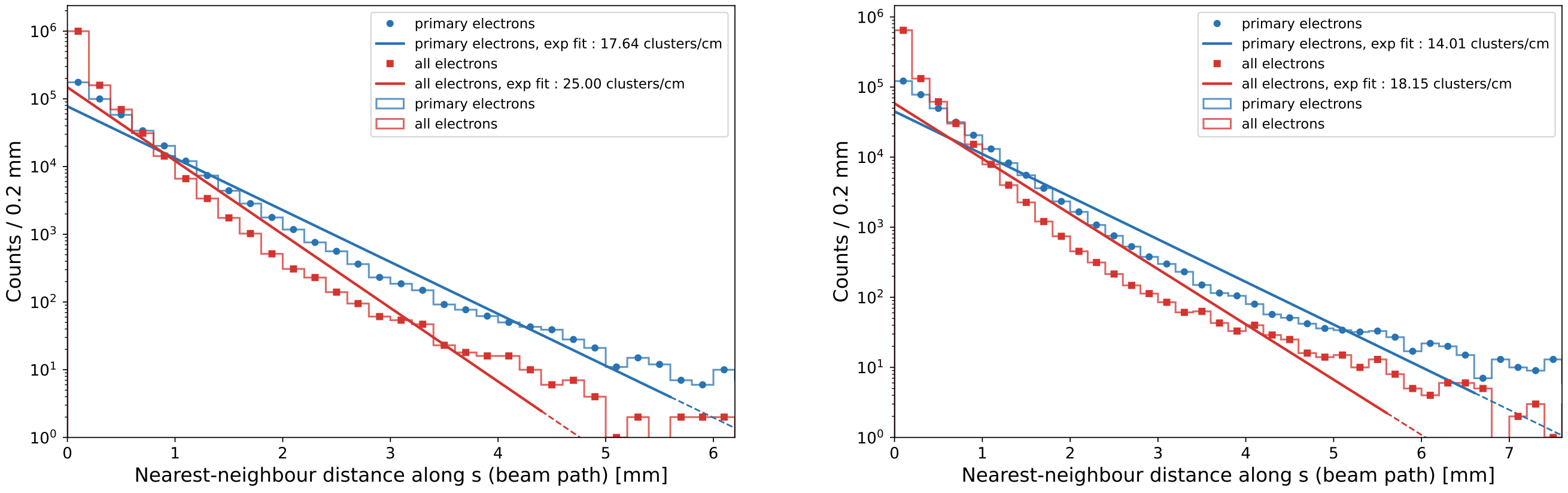

    \centering
    \begin{subfigure}[b]{0.49\textwidth}
        \centering
        \includegraphics[page=2, trim=0 0pt 0 75pt, clip, width=\textwidth]{images/monte_carlo/simulation_hit_characterisation/10tilt0rot_21mm_thinned_cluster_stats.pdf}
        \caption{}
        \label{fig:s_spacing_truth}
    \end{subfigure}
    \hfill
    \begin{subfigure}[b]{0.49\textwidth}
        \centering
        \includegraphics[page=8, trim=0 0pt 0 75pt, clip, width=\textwidth]{images/monte_carlo/simulation_hit_characterisation/10tilt0rot_21mm_thinned_cluster_stats.pdf}
        \caption{}
        \label{fig:s_spacing_processed}
    \end{subfigure}
    \caption{\footnotesize Nearest-neighbour distance along the beam trajectory $s$ for the $[10^\circ,0^\circ]$ configuration, for (a) the simulation truth dataset and (b) the processed simulation dataset after application of the known detector response and multiplicity matching. Each panel shows primary electrons only (blue circles) and all electrons produced along the track (red squares), together with the corresponding histograms with a bin width of $0.2\,\mathrm{mm}$. Solid lines show exponential fits over $0.4$--$3\,\mathrm{mm}$ in (a) and $0.4$--$4\,\mathrm{mm}$ in (b); dashed lines show their extrapolations. The two panels use different horizontal ranges.}
    \label{fig:s_spacing}
\end{figure}

Fig.~\ref{fig:s_spacing} compares consecutive-electron spacings along the tilt-corrected beam coordinate $s$ for $[10^\circ,0^\circ]$. Primary electrons contribute one position per primary ionisation; the all-electron sample also includes secondary ionisation. Independent, uniformly distributed ionisations give an exponential spacing distribution, so exponential fits provide a useful diagnostic rather than a complete description of either sample.

At truth level, Fig.~\ref{fig:s_spacing}(\subref{fig:s_spacing_truth})
shows that fits over $0.4$--$3\,\mathrm{mm}$ give slopes of
$17.64\,\mathrm{cm^{-1}}$ for primary electrons and
$25.00\,\mathrm{cm^{-1}}$ for all electrons. Primary spacings
approximately follow the exponential in this range. The all-electron
distribution has a pronounced short-distance excess from secondary
electrons produced near their parent ionisation. Ionisation by energetic
delta-electrons may also divide longer gaps, contributing to the steeper
all-electron slope. After the known response and multiplicity matching,
Fig.~\ref{fig:s_spacing}(\subref{fig:s_spacing_processed}) gives slopes of
$14.01\,\mathrm{cm^{-1}}$ and $18.15\,\mathrm{cm^{-1}}$,
respectively, for fits over $0.4$--$4\,\mathrm{mm}$. The corresponding
characteristic spacings increase from approximately $0.57$ to
$0.71\,\mathrm{mm}$ for primary electrons and from $0.40$ to
$0.55\,\mathrm{mm}$ for all electrons, consistent with the removal of
electrons by the detector response and multiplicity matching. The
short-spacing excess persists after processing, while both distributions
also exhibit a high-spacing tail above the extrapolated exponential.

The simulated electrons surviving diffusion, FEA, and multiplicity matching are converted into reconstructed longitudinal coordinates using the ToA response. A shutter duration of $17.25\,\mu\mathrm{s}$ and a Timepix clock period of $12.5\,\mathrm{ns}$ are applied before the arrival time is quantised to an integer clock count. The trigger-delay correction is $50\,\mathrm{ns}$ for the $21\,\mathrm{mm}$ measurements and $33\,\mathrm{ns}$ for the $5\,\mathrm{mm}$ measurement. The reconstructed drift coordinate is obtained from
\begin{equation}
z_{\mathrm{rec}} =
v_{\mathrm{d}}
\left(
t_{\mathrm{shutter}}
-n_{\mathrm{ToA}}T_{\mathrm{clk}}
+t_{\mathrm{trigger}}
\right),
\label{eq:toa_response}
\end{equation}
where $n_{\mathrm{ToA}}$ is the integer ToA clock count. The transverse coordinates are digitised to the centres of the physical Timepix pixel columns with pitch $p=55\,\mu\mathrm{m}$. This introduces a uniform quantisation uncertainty of $\sigma_{\mathrm{pix}}=p/\sqrt{12}\approx15.9\,\mu\mathrm{m}$, which is included in the transverse measurement covariance of the Kalman fit. At this stage, the simulated ionisation positions have been propagated through the principal gas-transport and detector-response effects and converted into pixel indices and ToA clock counts. The resulting dataset is processed using the same clustering, delta-electron rejection, and Kalman-filter reconstruction procedures as the test beam data.
\subsection{Probabilistic local-$z$ response}
\label{sec:local_z_response}
 Although the known response reproduces the main features of the reconstructed test beam distributions in the transverse directions, it does not fully reproduce the reconstructed longitudinal hit structure. Event-by-event studies of the reconstructed hit coordinates reveal differences in the local variation of the hit density and in the gap structure along the drift direction. This behaviour is quantified by comparing the fraction of empty and low-occupancy longitudinal bins, the hit-cluster spans, the $z$-span of gaps between clusters, and the correlation of hit counts between longitudinally separated $1\,\mathrm{mm}$-wide bins.

\begin{figure}[htbp]
    \centering
    \includegraphics[width=0.62\columnwidth]
        {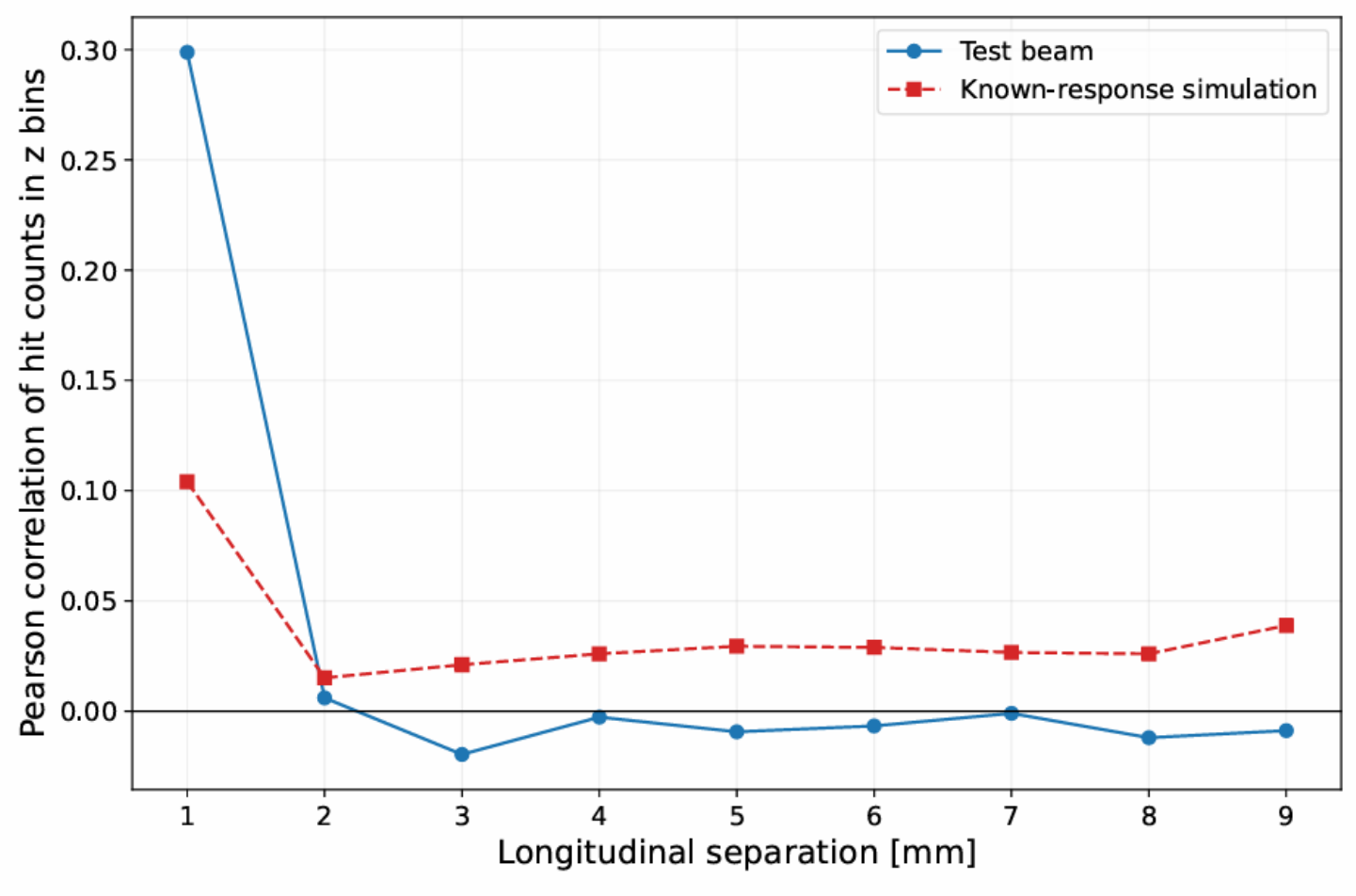}
    \caption{\footnotesize Longitudinal hit-pattern correlation for accepted tracks in the $[10^\circ,30^\circ]$ configuration. Each point gives the Pearson correlation between the hit counts in two $1\,\mathrm{mm}$-wide $z$ bins separated by the indicated longitudinal distance. Statistical uncertainties on the correlation coefficients are not shown.}
    \label{fig:longitudinal_hit_pattern_correlation}
\end{figure}

Fig.~\ref{fig:longitudinal_hit_pattern_correlation} shows a representative example for the $[10^\circ,30^\circ]$ configuration. The largest difference occurs at a separation of $1\,\mathrm{mm}$, where the correlation between hit counts in adjacent bins is substantially stronger in the test beam data than in the known-response simulation. At separations of $2\,\mathrm{mm}$ and above, both correlations become small, although the simulation retains a weak positive correlation at several separations. The test beam data therefore exhibit a short-range longitudinal hit-density structure that is not fully reproduced by the known-response simulation. Since the simulated multiplicity has already been matched to the test beam distribution, the remaining discrepancy concerns the local longitudinal arrangement of the recorded hits.

To reproduce this local longitudinal behaviour, a probabilistic local-$z$ response is applied after the known detector response and before clustering and track reconstruction. The model redistributes the spacings between neighbouring recorded hits, producing locally sparse and dense regions along the drift coordinate. It preserves the number of hits, the event multiplicity, the transverse hit coordinates, and the total longitudinal span.
For an event with ordered hit coordinates,
\begin{equation}
    z_1\leq z_2\leq\cdots\leq z_N,
    \qquad
    d_i=z_{i+1}-z_i,
\end{equation}
the consecutive longitudinal spacings are modified according to
\begin{equation}
    q_i=\exp(su_i),
    \qquad
    d_i^{\prime}
    =
    d_iq_i
    \frac{\sum_jd_j}{\sum_jd_jq_j}.
    \label{eq:local_z_spacing}
\end{equation}
The factor $q_i$ scales the $i$th spacing, while the final normalisation preserves the original total $z$ span. Since every $q_i$ is positive, the hit ordering is retained. The first and last longitudinal positions, the total number of hits, and all $x$ and $y$ coordinates remain unchanged.

The dimensionless parameter $s$ controls the strength of the local spacing variation. For each neighbouring-hit spacing, the latent variable $u_i$ determines whether that spacing is stretched or compressed. The position of the $i$th spacing is represented by its midpoint,
\begin{equation}
    c_i=\frac{z_i+z_{i+1}}{2}.
\end{equation}
Rather than being drawn independently, the variables $u_i$ form a correlated sequence along the track:
\begin{equation}
    u_i =
    \rho_i u_{i-1}
    +
    \sqrt{1-\rho_i^2}\,\varepsilon_i,
    \qquad
    \rho_i =
    \exp\left(
    -\frac{c_i-c_{i-1}}{\lambda}
    \right),
    \qquad
    \varepsilon_i\sim\mathcal{N}(0,1),
    \label{eq:local_z_field}
\end{equation}
starting from $u_1\sim\mathcal{N}(0,1)$, with the variables $\varepsilon_i$ drawn independently. For two spacings, this construction gives
\begin{equation}
    \operatorname{Corr}(u_i,u_j)
    =
    \exp\left(
    -\frac{|c_i-c_j|}{\lambda}
    \right).
    \label{eq:local_z_correlation}
\end{equation}
The parameter $\lambda$ therefore sets the longitudinal distance over which
neighbouring spacings tend to stretch or compress together, producing local
regions of lower or higher hit density.

For each reconstructed observable, the difference between the simulated
histogram $S_i$ and the scaled test beam histogram $R_i$ is quantified using
the symmetric metric
\begin{equation}
    \frac{\chi_{\mathrm{sym}}^2}{\mathrm{ndf}}
    =
    \frac{1}{N_{\mathrm{used}}-1}
    \sum_{i\in\mathcal{I}}
    \frac{(S_i-R_i)^2}{S_i+R_i},
    \label{eq:symmetric_chi2}
\end{equation}
where $\mathcal{I}$ contains the bins satisfying $S_i+R_i>5$, and
$N_{\mathrm{used}}$ is the number of retained bins. Before evaluating this
metric, the test beam histogram is scaled to the total number of simulated
entries. The metric is used to rank the tested response parameters and is not
interpreted as a formal goodness-of-fit probability.

A global score is defined as the equal-weight sum of the
configuration-averaged symmetric histogram scores for the four reconstructed
observables:
\begin{equation}
    \mathcal{S}
    =
    \left\langle
    \frac{\chi_{\mathrm{sym},\alpha}^{2}}{\mathrm{ndf}}
    \right\rangle
    +
    \left\langle
    \frac{\chi_{\mathrm{sym},\beta}^{2}}{\mathrm{ndf}}
    \right\rangle
    +
    \left\langle
    \frac{\chi_{\mathrm{sym},r_x}^{2}}{\mathrm{ndf}}
    \right\rangle
    +
    \left\langle
    \frac{\chi_{\mathrm{sym},r_y}^{2}}{\mathrm{ndf}}
    \right\rangle,
    \label{eq:global_score}
\end{equation}
where the brackets denote averages over the six detector configurations.

A scan over $s$ and $\lambda$ was first performed over a coarse grid to
identify the relevant parameter region. The scan was subsequently extended to
$s=0.150$--$0.350$ in steps of $0.025$ and
$\lambda=1.0$--$2.5\,\mathrm{mm}$ in steps of $0.1\,\mathrm{mm}$.
For each $(s,\lambda)$ candidate, the spacing redistribution was applied to
all known-response simulated events. The resulting events were then processed
through clustering and Kalman-filter track reconstruction, after which the
global score $\mathcal{S}$ was evaluated. The accepted-track longitudinal
diagnostics were also evaluated for each candidate as a separate check of the
resulting longitudinal hit structure, but were not included in the global
score or imposed as a hard selection requirement. The lowest four-observable
score was obtained for
\begin{equation}
    s=0.275,
    \qquad
    \lambda=1.8\,\mathrm{mm}.
\end{equation}

Applying the selected spacing redistribution to every simulated event
improved the longitudinal hit structure but broadened several reconstructed
track-observable distributions too strongly, most notably the $yz$ angular
width $\sigma_\beta$ in several configurations. This indicated that the
redistributed response should be applied only to a fraction of the simulated
event population.
With $s$ and $\lambda$ fixed at their selected values, an event fraction
$f_{\mathrm{gap}}$ was introduced. Each recorded simulated event receives the
local-$z$ spacing redistribution with probability $f_{\mathrm{gap}}$;
otherwise, its longitudinal hit coordinates remain unchanged. Clustering and
track reconstruction are subsequently applied to the resulting event sample.
The value of $f_{\mathrm{gap}}$ is determined from four accepted-track
longitudinal descriptors: the empty-bin fraction, the low-occupancy-bin
fraction, the largest gap relative to the mean gap, and the fraction of gaps
larger than twice the mean gap. For each descriptor, the simulated
distribution for a tested value of $f_{\mathrm{gap}}$ is represented by
\begin{equation}
    H_{\mathrm{mix}}(f_{\mathrm{gap}})
    =
    \left(1-f_{\mathrm{gap}}\right)H_{\mathrm{baseline}}
    +
    f_{\mathrm{gap}}H_{\mathrm{redistributed}},
    \label{eq:gap_fraction_mixture}
\end{equation}
where $H_{\mathrm{baseline}}$ and $H_{\mathrm{redistributed}}$ are the
accepted-track distributions obtained from the unmodified and fully
redistributed simulated samples, respectively. Each mixture distribution is
compared with the corresponding test beam distribution using the
total-variation distance. The selected value of $f_{\mathrm{gap}}$ minimises
the average total-variation distance over the four descriptors and the six
detector configurations.

A scan of $f_{\mathrm{gap}}$ over the unit interval gives
\begin{equation}
    f_{\mathrm{gap}}=0.43.
\end{equation}
The statistical uncertainty is evaluated by bootstrap-resampling the accepted
test beam tracks and repeating the minimisation, while the baseline and fully
redistributed simulation templates are kept fixed. The resulting bootstrap
$68\%$ interval is
\begin{equation}
    f_{\mathrm{gap}}\in[0.35,0.46].
\end{equation}
The local-spacing redistribution is therefore applied to approximately
$43\%$ of simulated events in the selected response model.
\subsection{Effective-smearing calibration}
\label{sec:effective_smearing}
 Although the local-$z$ response improves the longitudinal hit structure of the simulated tracks, residual differences remain in the reconstructed angular and spatial-residual distributions. A common effective-smearing model is therefore introduced to account collectively for response effects that are not fully captured by the available detector model. This is an empirical reconstruction-level response component applied in addition to the physical gas diffusion described above.
The transverse coordinates are assigned the same additional smearing width because the response is expected to be approximately symmetric in the $x$--$y$ plane. The reconstructed $z$ coordinate is inferred from the electron arrival time and is affected by the drift and ToA response, so an additional drift-dependent contribution is included. One relevant ToA effect is timewalk: signals of different amplitudes cross a fixed discriminator threshold at different times, shifting the inferred longitudinal coordinate. The effective-smearing model below is phenomenological and is not a calibrated amplitude-dependent timewalk correction:
\begin{equation}
   \sigma_{x,y}=\sigma_0,
   \qquad
   \sigma_z(z)=
   \sqrt{\sigma_0^2+\left(k_z z\right)^2}.
   \label{eq:additional_smearing}
\end{equation}
Here $\sigma_0$ is given in $\mu\mathrm{m}$, $z$ in mm, and $k_z$ is dimensionless. In this parametrisation, $\sigma_0$ gives the common coordinate-smearing scale, while $k_z$ describes any additional broadening of the reconstructed longitudinal coordinate with drift distance. The effective smearing is applied after the known detector response and the local-$z$ response, before clustering and Kalman-filter track reconstruction.

The parameters are determined through a simultaneous global optimisation of the reconstructed $\alpha$, $\beta$, $r_x$, and $r_y$ distributions across the six retained $21\,\mathrm{mm}$ configurations. The evaluated grid covers
\begin{equation}
    \sigma_0=20,22,\ldots,70\,\mu\mathrm{m},
    \qquad
    k_z=0,0.005,\ldots,0.050.
\end{equation}
For each configuration, a fixed and reproducible sample of 5000 simulated events is used throughout the calibration. Before evaluating the histogram metric, the test beam histograms are scaled to the corresponding number of simulated entries. The simulated angular distributions are re-centred on the corresponding reference angular means, while the simulated residual distributions are re-centred on the measured test beam residual means. Fixed differences in the distribution centres therefore do not influence the effective-smearing calibration.

The minimum of the scan is obtained for
\begin{equation}
    \sigma_0=58\,\mu\mathrm{m},
    \qquad
    k_z=0.045,
\end{equation}
with a global score of $\mathcal S=25.192$. The statistical uncertainty of the calibration is evaluated using $1000$ bootstrap replicas, giving the $68\%$ intervals
\begin{equation}
    \sigma_0\in[56,58]\,\mu\mathrm{m},
    \qquad
    k_z\in[0.040,0.045].
\end{equation}
The selected response parameters are summarised in Table~\ref{tab:response_model_parameters}. The response model calibrated with the $21\,\mathrm{mm}$ data is subsequently applied without re-tuning to the independent $5\,\mathrm{mm}$ measurement, providing a separate test of the response prescription at a different drift length and track orientation.
\begin{table}[!ht]
\centering
\small
\caption{Parameters of the final empirical GridPix response model calibrated with the six retained $21\,\mathrm{mm}$ configurations. The last column gives the bootstrap $68\%$ interval from the corresponding calibration stage.}
\label{tab:response_model_parameters}
\renewcommand{\arraystretch}{1.15}
\begin{tabular}{cccc}
\hline
\textbf{Response stage} & \textbf{Parameter} &
\textbf{Selected value} & \textbf{Bootstrap $68\%$ interval} \\
\hline
Local-$z$ spacing response
& $s$
& $0.275$
& $[0.275,\,0.325]$ \\
Local-$z$ spacing response
& $\lambda$
& $1.8\,\mathrm{mm}$
& $[1.1,\,1.8]\,\mathrm{mm}$ \\
Probabilistic event mixture
& $f_{\mathrm{gap}}$
& $0.43$
& $[0.35,\,0.46]$ \\
Effective smearing
& $\sigma_0$
& $58\,\mu\mathrm{m}$
& $[56,\,58]\,\mu\mathrm{m}$\\
Effective smearing
& $k_z$
& $0.045$
& $[0.040,\,0.045]$ \\
\hline
\end{tabular}
\end{table}

\subsection{Uncertainty estimation}
\label{sec:uncertainty_estimation}
Data and simulation uncertainties are evaluated separately. For each width $W\in\{\sigma_\alpha,\sigma_\beta,\sigma_{r_x},\sigma_{r_y}\}$, the data uncertainty is the largest shift from the central $10\times10\,\mathrm{mm}^2$ result when using central $9\times9$ and $11\times11\,\mathrm{mm}^2$ regions:
\begin{equation}
\delta W_{\mathrm{area}} =
\max\left(
\left|W_{9\times9}-W_{10\times10}\right|,
\left|W_{11\times11}-W_{10\times10}\right|
\right).
\label{eq:area_systematic}
\end{equation}
The area variation probes sensitivity to edge fields, beam position, and modest changes in track containment. It is assigned to the central $10\times10\,\mathrm{mm}^2$ result.

Response-calibration uncertainties are obtained by multinomial resampling of the data histogram counts at fixed entry totals. Each replica selects the evaluated $(s,\lambda)$ or $(\sigma_0,k_z)$ grid point with the lowest $\mathcal S$; all local-$z$ candidates remain eligible, with longitudinal diagnostics used separately. For $f_{\mathrm{gap}}$, complete accepted data tracks are resampled and the mixture fit repeated with fixed simulation templates, preserving correlations among fragmentation descriptors.
The bootstrap spread measures sensitivity to the available calibration data on the evaluated parameter grids. Each stage uses the selected central values of the preceding stages, rather than a simultaneous refit of the full response. Selected values and $68\%$ intervals are listed in Table~\ref{tab:response_model_parameters}.

Local-$z$ variations, including $f_{\mathrm{gap}}$, and effective-smearing variations are propagated through reconstruction. With no joint covariance available, their contributions are treated as independent:
\begin{equation}
\delta_{\mathrm{staged},\pm} =
\sqrt{
\left(\delta_{\mathrm{local}\text{-}z,\pm}\right)^2+
\left(\delta_{\mathrm{eff},\pm}\right)^2
}.
\label{eq:staged_response_uncertainty}
\end{equation}
The two terms describe variations from $(s,\lambda,f_{\mathrm{gap}})$ and $(\sigma_0,k_z)$; upper and lower shifts are propagated separately.

With empirical parameters fixed, diffusion is varied one coefficient at a time:
\begin{equation}
D_{\mathrm T}=\{80.5,93.3,106.0\},\qquad
D_{\mathrm L}=\{107.4,117.4,127.4\}\;\mathrm{cm^2\,s^{-1}}.
\label{eq:DT_systematic_variations}
\end{equation}
Each variation is propagated through the detector response and complete reconstruction without re-optimising the empirical parameters. The middle values are the central coefficients. For $D_{\mathrm T}$, the
difference between the calibrated central value and the nominal Magboltz
prediction, $12.7\,\mathrm{cm^2\,s^{-1}}$, is applied symmetrically around
the calibrated value. The nominal Magboltz prediction therefore corresponds
to the upper variation. This interval is a systematic variation and not a
statistical confidence interval. For $D_{\mathrm L}$, the nominal Magboltz
prediction remains the central value, and the lower and upper variations
correspond to its propagated $8.5\%$ relative uncertainty. 

For $j=\mathrm T,\mathrm L$, let $\Delta_{j\pm}=W(D_j^\pm)-W_0$. The signed envelopes are
\begin{equation}
\delta_{D_j}^{+}=\max(0,\Delta_{j-},\Delta_{j+}),\qquad
\delta_{D_j}^{-}=\max(0,-\Delta_{j-},-\Delta_{j+}).
\label{eq:diffusion_width_shifts}
\end{equation}
Here $W_0$ is the width from the central simulation. Taking the largest positive and negative shifts separately accommodates asymmetric and non-monotonic changes; it does not assume that increasing diffusion necessarily increases every reconstructed width. Assuming independent contributions, the final simulation uncertainty is
\begin{equation}
\delta_{\mathrm{MC},\pm}^2=\delta_{\mathrm{staged},\pm}^2+(\delta_{D_{\mathrm T}}^\pm)^2+(\delta_{D_{\mathrm L}}^\pm)^2.
\label{eq:total_simulation_uncertainty}
\end{equation}

These are systematic envelopes rather than a joint confidence region. Correlations between calibration stages and diffusion, and uncertainties from Geant4 physics, alignment, other transport effects, and the local-$z$ functional form are excluded. Data and simulation errors are shown separately.

\section{Results and Conclusions}
\label{sec:Results}
\begin{figure*}[htbp]
    \centering
    \includegraphics[width=1\textwidth]
        {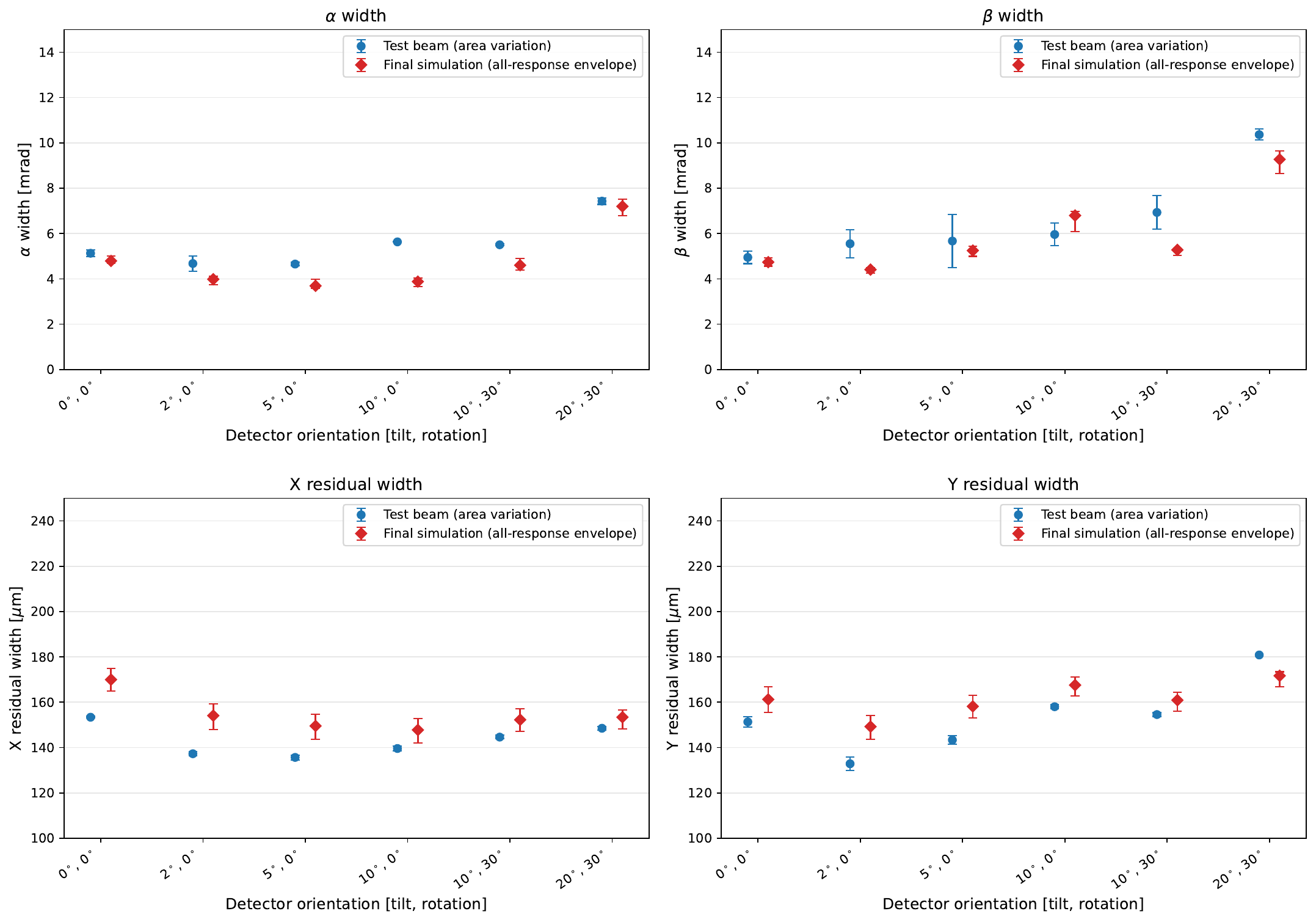}
    \caption{\footnotesize Comparison of the reconstructed test beam widths and
the selected response-model predictions for the six analysed
$21\,\mathrm{mm}$ configurations. The test beam error bars represent the
variation obtained from the central $9\times9\,\mathrm{mm}^2$ and
$11\times11\,\mathrm{mm}^2$ fiducial regions relative to the central
$10\times10\,\mathrm{mm}^2$ result. The simulation error bars combine the
staged response-calibration uncertainty and the propagated transverse- and
longitudinal-diffusion variations.}
    \label{fig:final_21mm_comparison}
\end{figure*}

 As described in Section~\ref{sec:Detector_Response_Model}, applying
the known detector response alone to the Geant4 truth data does not reproduce
the longitudinal hit pattern and the reconstructed angular and
spatial-residual widths simultaneously. The probabilistic local-$z$ response
accounts for the observed local variations in hit density along the drift
direction, while the common effective-smearing term brings the reconstructed
observables into closer simultaneous agreement with the test beam data.
Fig.~\ref{fig:final_21mm_comparison} compares the resulting response model
with the reconstructed widths for all six $21\,\mathrm{mm}$ configurations.
The test beam and simulation uncertainties shown in
Fig.~\ref{fig:final_21mm_comparison} are evaluated as described in
Section~\ref{sec:uncertainty_estimation}. All twelve spatial-residual widths
agree with the test beam values within approximately $12\%$, while ten of the
twelve angular widths agree within approximately $20\%$. The largest
remaining difference occurs for the $[10^\circ,0^\circ]$ configuration, where
the simulated $\alpha$ width is approximately $31\%$ smaller than the test
beam value. For the $[10^\circ,30^\circ]$ configuration, the simulated
$\beta$ width is approximately $24\%$ smaller than the measurement.

\begin{figure*}[htbp]
    \centering
    \includegraphics[width=1\textwidth]
        {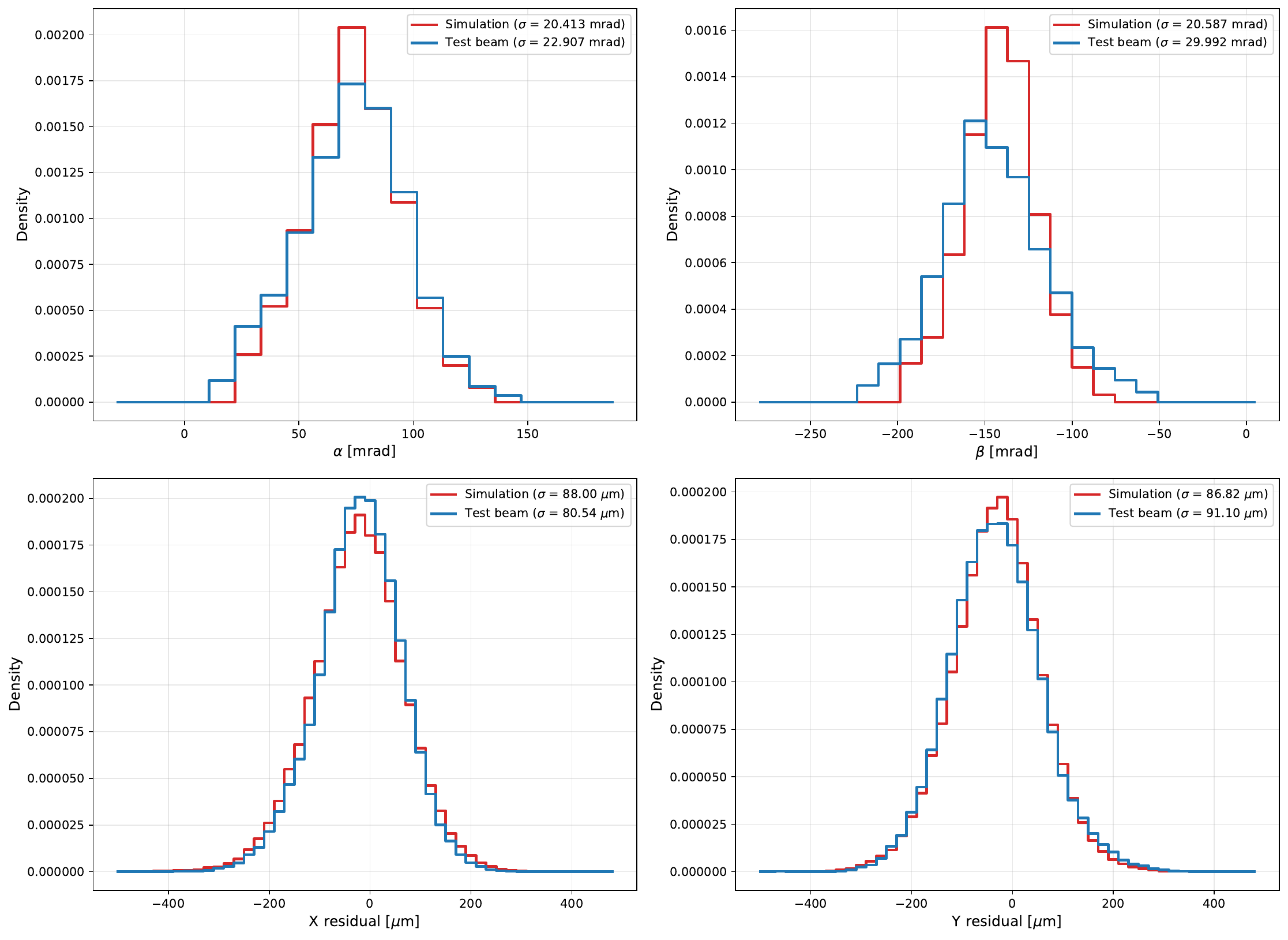}
    \caption{\footnotesize Comparison of the reconstructed $\alpha$, $\beta$,
X-residual, and Y-residual distributions for the independent
$5\,\mathrm{mm}$ $[10^\circ,20^\circ]$ measurement and the simulation obtained
by applying the response model calibrated with the $21\,\mathrm{mm}$ data
without re-tuning.}
    \label{fig:validation_5mm}
\end{figure*}

 The response calibrated with the $21\,\mathrm{mm}$ data was finally
applied, without re-tuning, to the independent $5\,\mathrm{mm}$
$[10^\circ,20^\circ]$ measurement. As shown in
Fig.~\ref{fig:validation_5mm}, the simulated $\alpha$ and $\beta$ widths are
approximately $11\%$ and $31\%$ smaller than the corresponding test beam
values, respectively. In contrast, the simulated X-residual width is
approximately $9\%$ larger, while the Y-residual width is approximately
$5\%$ smaller.

A possible contribution to the larger discrepancy observed for the angular widths is indicated by the hit-multiplicity distributions of the selected events. The mean multiplicities differ by approximately $5$ hits per event, with $27.73$ hits per event in the test beam data and $22.71$ in the simulation. Despite their similar low-multiplicity peaks, the test beam distribution has a larger fraction of events away from the peak, resulting in a standard deviation of $29.69$ hits compared with $14.88$ hits in the simulation. This difference in the multiplicity distributions can modify the clustering and the effective longitudinal hit lever arm used by the Kalman fit, to which the angular observables are particularly sensitive. The transverse residuals are more local observables and are consequently less affected. The multiplicity difference therefore provides a plausible contribution to the remaining angular discrepancy, although the present comparison does not establish it as its sole origin. This comparison tests the application of the fixed $21\,\mathrm{mm}$ response prescription to one measurement at a different drift length and track orientation.

The present study should be regarded as a preliminary development of the
GridPix response model. In several configurations, the beam was offset from
the chip centre, while the mechanical tilt--rotation setting had limited
precision and the electric field behaviour near the chip edges could not be
monitored directly. The Timepix chip was operated in ToA mode, so no
simultaneous Time-over-Threshold (ToT) measurement of the signal amplitude was
available for the same pixel hit. Avalanche-amplitude fluctuations cause
timewalk because the threshold-crossing delay depends on the signal amplitude.
Combined timing and charge information would allow this dependence to be
studied and corrected. Such a correction of the reconstructed longitudinal
coordinate could not be investigated with the present data. A future
measurement with improved beam alignment, a more precise tilt--rotation
system, and controlled field conditions would allow the angular response to
be studied more reliably. The addition of an external beam telescope would provide an independent
reference trajectory and allow the intrinsic spatial resolution to be
estimated, which cannot be achieved from the internal track-fit residuals
alone.

This work provides a single-layer tracklet characterisation and an empirical Geant4-to-readout response for the measured conditions. The model is a starting point for compact tracking studies, including vertex association near an emulsion detector or in a gaseous active target. Establishing performance in those applications requires stacked-layer measurements, further drift lengths and gas mixtures, and track momenta and multiplicities representative of the intended experiment.

\begin{acknowledgments}
The authors thank the ELSA accelerator team at the University of Bonn, in
particular Dr.\ Frank Frommberger, for their support during the test beam
measurements.
\end{acknowledgments}
\bibliographystyle{apsrev4-2}
\bibliography{references}
\end{document}